\documentclass[twoside,11pt,fleqn]{article}
\usepackage{epsfig,psfig,amssymb,amstext,ijbc}

\newcommand{\figdir}{./}

\newcommand{\equ}{Eq.}
\newcommand{\fig}{Fig.}
\newcommand{\figs}{Figs.}

\newcommand{\rem}[1]{}

\newtheorem{platz}{{Fig.}} 
\newcommand{\FIGo}[3]{\begin{figure}%
#3%
\caption[]{\footnotesize #2}%
\label{#1}%
\end{figure}}

\rem{
\newcommand{\FIGo}[3]{
  \begin{center}
    \begin{minipage}{16.0cm}
    \begin{picture}(16.0,1)
      \put(0,0){\framebox(16.0,1){\ }}
    \end{picture}
      \begin{footnotesize}
        \begin{platz}   \label{#1}
          {\rm #2}
        \end{platz}
      \end{footnotesize}
    \end{minipage}
  \end{center}
  }
}

\rem{
\newcommand{\FIGo}[3]{%
\marginpar{\begin{platz} \label{#1} ~ \end{platz} \vspace*{1.5ex} }
}

}

\newcommand{\N}{{\mathbb N}}

   \def\bege{\begin{equation}}
   \def\ende{\end{equation}}
   \def\bega{\begin{eqnarray}}
   \def\enda{\end{eqnarray}}
   \def\began{\begin{eqnarray*}}
   \def\endan{\end{eqnarray*}}

\newcommand{\eqref}[1]{(\ref{#1})}

\newcommand{\arn}{Arnol$'$d }
\newcommand{\ve}{\varepsilon}
\newcommand{\vt}{\vartheta}

\title{\vspace*{-2cm}
\Huge{\bf Multistability and nonsmooth bifurcations in the quasiperiodically forced circle map}
}
 
\author{
        {\sc Hinke Osinga} \\
	School of Mathematical Sciences, \\ 
	University of Exeter, 
	Exeter EX4 4QE, UK \\
        {\small h.m.osinga@exeter.ac.uk}
\\
        {\sc Jan Wiersig} \\
	Max-Planck-Institut 
	f\"ur Physik komplexer Systeme, \\
	D-01187 Dresden, Germany \\
	{\small jwiersig@mpipks-dresden.mpg.de}
\\
	{\sc Paul Glendinning} \\
        Department of Mathematics, \\
	UMIST, 
	Manchester M60 1QD, UK \\
        {\small p.a.glendinning@umist.ac.uk}
\\
        {\sc Ulrike Feudel} \\
        Institut f\"ur Physik, \\
        Universit\"at Potsdam, PF 601553,  
        D-14415 Potsdam, Germany \\
        {\small ulrike@agnld.uni-potsdam.de}
       }
 
\begin{document}

\begin{titlepage}
\maketitle
 

\begin{abstract}
\noindent
It is well-known that the dynamics of the \arn circle map is 
phase-locked in regions of the parameter space called \arn
tongues. If the map is invertible, the only possible dynamics is
either quasiperiodic motion, or phase-locked behavior with a unique
attracting periodic orbit. Under the influence of quasiperiodic
forcing the dynamics of the map changes dramatically. Inside the \arn
tongues open regions of multistability exist, and the parameter
dependency of the dynamics becomes rather complex. This paper
discusses the bifurcation structure inside the \arn tongue with
zero rotation number and includes a study of nonsmooth
bifurcations that happen for large nonlinearity in the region with 
strange nonchaotic attractors.
\end{abstract}
\end{titlepage}
 
\pagestyle{myheadings}

 
\section{Introduction}
\label{sec:intro}

The \arn circle map \cite{Arnold65} 
\begin{equation}
\label{eq:arn}
x_{n+1} = x_n + \Omega + \frac{K}{2\pi} \sin(2 \pi x_n) \;\; {\rm mod \; 1}
\end{equation}
is one of the paradigms for studying
properties of nonlinear dynamical systems, both because it is a very
simple map, and because of its great physical relevance (see,
e.g.~\cite{BBJ85}). Using the circle map one can model the
structure of phase-lockings (devil's staircase) of a periodically
forced nonlinear oscillator \cite{JBB83,JBB84} and 
the current-voltage characteristics of a driven Josephson junction
\cite{BBJ84}. The phase-locked regions of the \arn circle map form the
well-known \arn tongues \cite{Arnold83,Hall84}. If $|K|< 1$ there is a
unique periodic attractor with a particular rotation number in each tongue. 

In this paper we study the structure of the phase-locked regions of
the \arn circle map driven by a rigid rotation with an irrational
frequency. This system exhibits different kinds of dynamics, namely
quasiperiodic motions with two and three incommensurate frequencies,
chaotic attractors, and strange nonchaotic attractors
(SNAs). SNAs have a strange geometrical structure, but unlike chaotic 
attractors they do not exhibit a sensitive dependence to changes
in the initial conditions, i.e. their dynamics is not chaotic. They
have been found in many quasiperiodically forced systems \cite{GOPY84}, 
and
also in the quasiperiodically forced circle map
\cite{DGO89,FKP95}.

Previous investigations show that regions of bistability occur in
phase-locked regions of the quasiperiodically forced circle map
\cite{GFPS99} and the phase-locked regions change in shape depending
on the strength of the forcing. This change in shape is related 
to the emergence of
SNAs \cite{FGO97}. Based on these studies the aim of this paper is
twofold. We study regions in parameter space where more than two
attractors coexist (pockets of
multistability). Secondly, we discuss the relation between these
multistable regions and the appearance of strange nonchaotic
attractors. In our discussion of the way these attractors are created 
and destroyed we are led to a description of nonstandard (nonsmooth) 
bifurcations of the invariant curves. 
 
If the unforced 
circle map is modified by introducing additional nonlinearities,
coexisting attractors with the same rotation number can occur within
the phase-locked regions \cite{McGP96}. 
In our system, multistability within the phase-locked regions is
induced by the forcing rather than an additional nonlinear term. 
In quasiperiodically forced systems the coexisting attractors 
may be either invariant curves or SNAs
depending on the strength of the forcing. Our investigation
focuses on how these multistable regions appear and
disappear under variation of the system's parameters: the nonlinearity 
$K$ and the forcing amplitude $\varepsilon$. For the tongue with zero
rotation number the multistable regions open and close by {\it smooth}
saddle-node or pitchfork bifurcations of invariant curves if 
$K$ and $\varepsilon$ are small. For larger $K$ these saddle-node 
and pitchfork bifurcations become {\it nonsmooth}: instead of merging
uniformly (smooth bifurcation),  the relevant stable and unstable
invariant curves appear to collide only in a dense set of points.

The paper is organized as follows. Section \ref{sec:qfcm} recalls
important properties of the \arn circle map relevant for this study
and their changes under the influence of quasiperiodic forcing. In
particular we discuss the phase-locked region with zero rotation
number. Within this phase-locked region we find pockets of
multistability with a rather complex bifurcation structure which is
analyzed in Sec.~\ref{sec:maintongue}. Smooth and nonsmooth
saddle-node and pitchfork bifurcations, leading to coexisting
attractors, are studied in Sec.~\ref{sec:epsK} to get a better
understanding of the changes in the bifurcation structure depending on
the strength of nonlinearity and forcing. Furthermore, we investigate
the transition between smooth and nonsmooth bifurcations and its
implications to the dynamics of the system. In the full parameter space 
we find bifurcations of codimension two. In Sec.~\ref{sec:nscodim} we 
discuss a special codimension-2 point that involves only nonsmooth 
bifurcations. It turns out that the unfolding of this point is very 
different from the smooth analog. Finally, in
Sec.~\ref{sec:neighbourhood}, we briefly discuss phase-locked regions with
small, but finite, rotation number. We conclude this paper with a summary in
Sec.~\ref{sec:disc} and an Appendix with details on the numerical
computations. 
For readers with a black and white copy of this article we provide 
a supplementary website~\cite{OWGFwww00}.
 
 
\section{The Quasiperiodically Forced Circle Map}
\label{sec:qfcm}

The quasiperiodically forced circle map is a map on the torus with 
lift 
\begin{eqnarray}
\label{eq:mapx}
   x_{n+1} & = & x_n + \Omega + \frac{K}{2\pi} \sin{(2\pi x_n)} +
           \ve \sin{(2\pi \vartheta_n)} \ , \\
\label{eq:maptheta}
   \vartheta_{n+1} & = & \vartheta_n + \omega \;\; {\rm mod \; 1} ,
\end{eqnarray}
where $\vartheta_n$ and $x_n$ modulo 1 give the coordinates on the torus. 
The parameter
$\Omega$ is the phase shift, $K$ denotes the strength of nonlinearity 
($K > 0$), $\ve$ is the forcing amplitude, and the forcing frequency 
$\omega$ is irrational. Throughout this paper we choose to work with 
$\omega = (\sqrt{5} - 1) / 2$.

\subsection{The unforced system}

Let us recall the behavior of the unforced circle map \eqref{eq:arn}. 
The dynamics of this map can be either
periodic, quasiperiodic, or chaotic, depending on the parameters
$\Omega$ and $K$. The critical line $K = 1$ divides the parameter
space into two regions. If $K < 1$ the map is invertible and the
motion can only be periodic (phase-locked) or quasiperiodic. 
For $K > 1$ the map is
noninvertible and chaotic motion is possible. 

The rotation number is used to characterize the different kinds of
motion. It is defined as
\begin{equation}
\label{eq:rho}
        \rho(\Omega, K) = \lim_{N \to \infty} 
                               \frac{x_N - x_0}{N}, 
\end{equation}
where $x_N$ is the $N$th iterate of \eqref{eq:arn}, 
starting from $x_0$. It can be shown that $\rho(\Omega, K)$ does not
depend on $x_0$ if $K < 1$. If the rotation number is
rational, the attracting motion is periodic, otherwise it is quasiperiodic. For
$K < 1$ the parameter space is split into regions with rational
rotation number, the phase-locked regions or \arn tongues, and
regions with irrational rotation number corresponding to quasiperiodic
motion. For example, the main tongue, the phase-locked region with
zero rotation number, is bounded by the curves $\Omega = \pm
\Omega_0(K)$, where $\Omega_0(K) = K / 2\pi$. For any choice of $\Omega$ 
and $K$, with $|\Omega| < \Omega_0$ and $K < 1$, 
there are exactly two fixed points, one is
attracting and the other is repelling. At the boundary $|\Omega| =
\Omega_0$ the two fixed points are annihilated in a saddle-node
bifurcation. For other rotation numbers $\rho \neq 0$ the
$K$-dependency of the boundary $\Omega_{\rho}(K)$ is nonlinear, but it
is always a curve of saddle-node bifurcations.

\subsection{The forced system}
\label{sec:app1}

A variety of behavior is possible in the coupled maps 
(\ref{eq:mapx})--(\ref{eq:maptheta}).
The rotation number (\ref{eq:rho}) exists, but depends on $\ve$ and 
$\omega$ in addition to $\Omega$ and $K$, and the direct
analogs of the periodic and quasiperiodic motion of the
uncoupled \arn map are invariant curves (the
graph of a function $\vartheta \mapsto x(\vartheta)$) and motion which
is dense on the torus, respectively. In the former
case the rotation number is rationally related to $\omega$
($\rho = r_1+r_2\omega$ with $r_i$ rational, $i=1,2$) and
in the latter case there is no such rational relation. For both rationally 
and irrationally related rotation numbers 
strange nonchaotic attractors (SNAs) may also be possible.
An SNA has a strange geometric structure, that is, it can be viewed as the
graph of an everywhere discontinuous function $\vartheta \mapsto
x(\vartheta)$, but the dynamics on the attractor is not chaotic, because  
typical Lyapunov exponents  in the $x$--direction are negative 
(there is always a zero Lyapunov exponent in the $\vt$--direction).

If the rotation number is rationally related to $\omega$ then the
motion is said to be {\it phase-locked} and the regions of
parameter space in which the motion is phase-locked are analogous
to the \arn tongues of the unforced map. On the boundaries of the
phase-locked regions we expect to see saddle-node bifurcations. There 
is an additional complication in the forced maps \cite{FKP95} in 
that the saddle-node bifurcations may be smooth (two invariant 
curves converge uniformly from inside the phase-locked region) or
nonsmooth. In the smooth saddle-node bifurcation the nontrivial Lyapunov
exponent in the $x-$direction goes to zero at the bifurcation
point. In the nonsmooth saddle-node bifurcation the two invariant
curves appear to collide only on a dense 
set of points. Moreover, the typical nontrivial Lyapunov exponent
remains negative. These nonsmooth saddle-node bifurcations seem to 
be associated with the appearance of SNAs outside the phase-locked 
region \cite{DGO89,FKP95,Glendinning98}.

\def\figmaintongue{%
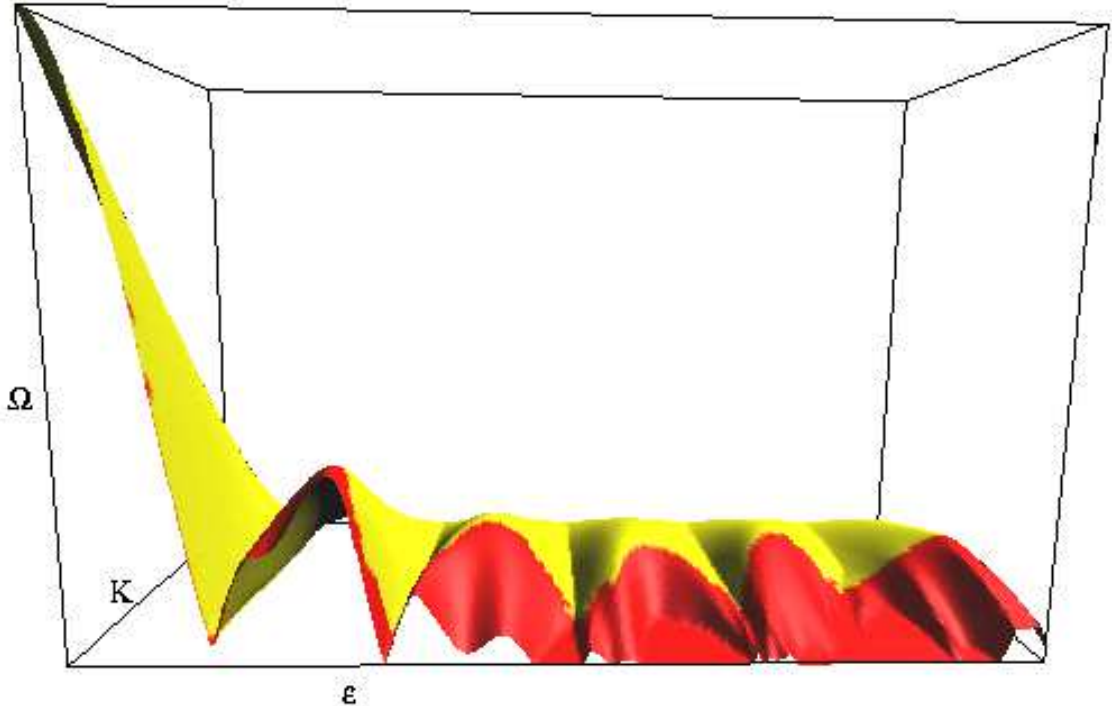
The boundary $|\Omega| = \Omega_0(\varepsilon, K)$ of the phase-locked 
region with zero rotation number; $\varepsilon \in [0,5]$ runs from
left to right, $K \in [0, 1]$ from back to front, and $|\Omega| \in
[0, 0.16]$ from bottom to top.  Pairs of invariant curves inside the 
phase-locked region ($|\Omega| < \Omega_0$) are annihilated at its
boundary leading either to three-frequency quasiperiodic motion
(yellow area) or SNAs (red area); see App.~\ref{app:sec22}.
}
\def\FIGmaintongue{ 
\centerline{\psfig{figure=\figdir/maintongue.eps,width=15.0cm,angle=0}}}
\FIGo{fig:maintongue}{\figmaintongue}{\FIGmaintongue}
It can be shown that one mechanism of the appearance of SNAs is 
related to changes in 
the shape of the phase-locked regions \cite{FGO97,GFPS99}. For
the unforced map the width of a phase-locked region increases
monotonically with increasing nonlinearity $K$; this is no longer the
case for positive forcing amplitude $\varepsilon$. Moreover, as 
Fig.~\ref{fig:maintongue} shows, for fixed $K$ 
the width of the phase-locked region oscillates as $\varepsilon$ increases. 
In particular, there are certain values  of $\varepsilon$ for which the
width of the phase-locked region becomes extremely
small. Unfortunately, using only numerical methods, we cannot decide
whether the region actually closes or not at those
$\varepsilon$-values. For more details on numerical computations we refer 
to App.~\ref{app:sec22}. 

For small fixed nonlinearity $K$ the boundary $\Omega_0(\varepsilon, K)$ of
the phase-locked region can be approximated by (the modulus of) a Bessel
function of order zero using 
first order perturbation theory~\cite{GFPS99}. Numerical simulations
also revealed regions of bistability in the vicinity of the zeroes of
the Bessel function, where the width of the phase-locked region is
very small. The bistability regions are bounded by saddle-node
bifurcations of invariant curves, which has been confirmed by second
order perturbation theory~\cite{GW99}.

The study in \cite{GFPS99,GW99} only applies for $K$ close to $0$. We
wish to study what happens to the phase-locked region with zero
rotation number for larger $K$. However, we restrict our
considerations to the invertible case $K < 1$, so that chaos is ruled
out. We find that the regions of bistability contain other regions
where even more attractors coexist. In the following we describe how
these regions appear and disappear as a parameter varies. We also study
smooth and nonsmooth bifurcations and make some remarks on the appearance 
of SNAs. 

The majority of the rest of this paper describes the results
of numerical simulations of Eqs.~(\ref{eq:mapx})--(\ref{eq:maptheta}). 
As such, the reader should
bear in mind that our conclusions are based on numerical observations
and may turn out to be misleading in places. We have made every effort
to avoid such problems (see the Appendix) and believe that the
phenomena reported are sufficiently interesting and mathematically
intractable to merit this numerical investigation, even if we
remain uncertain of some of the outcomes. The reader is encouraged to
maintain a healthy scepticism throughout.

 
\section{The Internal Structure of the Main Tongue for $K = 0.8$}
\label{sec:maintongue}

\begin{figure}[t]
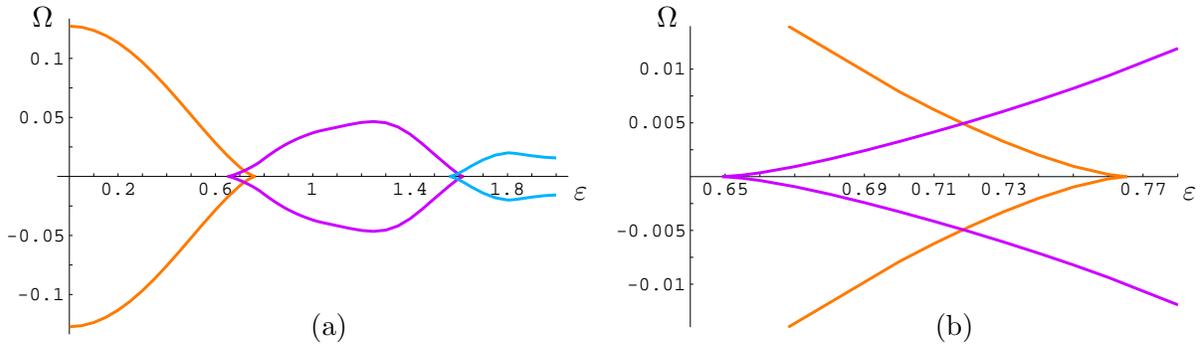

\begin{picture}(160,45)(0,0)
\put(0,0){
\psfig{file=\figdir/fullepsOmega.ps,width=75mm}
}
\put(5,41){$\Omega$}
\put(77,18){$\ve$}
\put(42,0){(a)}
\put(81,1){
\psfig{file=\figdir/fstoverlap.ps,width=75mm}
}
\put(88,41){$\Omega$}
\put(158,18){$\ve$}
\put(125,0){(b)}
\end{picture}
\caption{\footnotesize
The bifurcation structure for $K = 0.8$ of the phase-locked region with
zero rotation number (a) with detail of the first overlap (b).
}
\label{fig:epsOmega}
\end{figure}
In the simplest case the boundary $|\Omega| = \Omega_0(\ve, K)$ represents the
disappearance of two invariant curves, a stable and an unstable
one. However, for $|\Omega| < \Omega_0(\ve, K)$ more than two
invariant curves may exist that disappear before this boundary 
is crossed. Such pockets of multistability are found near local minima
of $\Omega_0(\ve, K)$, cf. the region of bistability predicted by 
perturbation analysis for small $K$ \cite{GFPS99,GW99}. 
For example, Fig.~\ref{fig:epsOmega}(a)
shows a cross-section of Fig.~\ref{fig:maintongue} at $K = 0.8$,
with both positive and negative sides of the boundary of the
phase-locked region. The outer boundary is the function $|\Omega| =
\Omega_0(\ve, 0.8)$. Extra curves are drawn marking the boundaries
of pockets of multistability, which is best seen in the enlargements
Figs.~\ref{fig:epsOmega}(b), \ref{fig:swallowtail} and
\ref{fig:subpf}. These pockets of multistability can be considered as
overlaps of different ``bubbles'' with the same rotation number 
as in Fig.~\ref{fig:epsOmega}(a). For
better visualization we have chosen different colors for bifurcations
of different pairs of invariant curves.

\subsection{Bifurcations for $K = 0.8$ in the first region of overlap}
\label{sec:easy}

\begin{figure}[t]
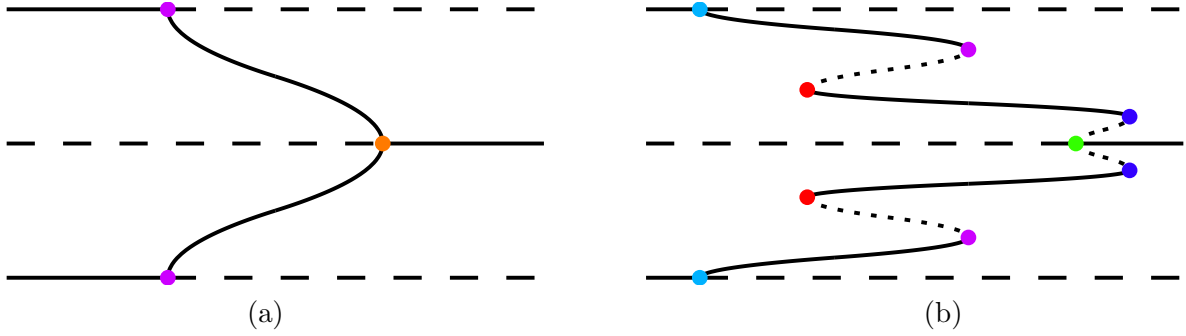

\begin{picture}(160,45)(0,0)
\put(0,5){
\psfig{file=\figdir/bifportr1.ps,width=75mm}
}
\put(35,0){(a)}
\put(85,5){
\psfig{file=\figdir/bifportr2.ps,width=75mm}
}
\put(125,0){(b)}
\end{picture}
\caption{\footnotesize
Sketch of the bifurcations along the line $\Omega = 0$ and $K = 0.8$
in the first (a) and second (b) overlap. Shown are invariant circles
(represented by one point) versus $\ve$. Closed curves represent
stable and dashed curves represent unstable circles. The colors
correspond with the colors of the bifurcation curves in
Figs. \protect\ref{fig:epsOmega}, \protect\ref{fig:swallowtail} and
\protect\ref{fig:subpf}.
}
\label{fig:bifportraits}
\end{figure}

The first overlap is enlarged in Fig.~\ref{fig:epsOmega}(b). The
orange and purple curves enclose a rhombus shaped region where two
attracting and two repelling invariant circles exist. This region of
bistability is bounded by curves of saddle-node bifurcations that end
in pitchfork bifurcations on the line $\Omega = 0$. 
Note that if $\Omega = 0$ the map has a symmetry 
($x \mapsto -x,\; \vt \mapsto \vt+1/2$) which implies that the 
rotation number in the $x$--direction is always zero and pitchfork 
bifurcations should be expected. 
The bifurcation sequence for $\Omega = 0$ is sketched in
Fig.~\ref{fig:bifportraits}(a) where each circle is represented as a
point and $\ve$ increases along the horizontal axis. The bottom and
top lines are identical, representing the modulo 1 computations.
The purple and orange dots are the pitchfork bifurcations that mark
the crossing of the purple and orange curves in
Fig.~\ref{fig:epsOmega}(b) along $\Omega = 0$, respectively.

\subsection{Bifurcations for $K = 0.8$ in the second region of overlap}
\label{sec:epsOmega}

\begin{figure}[t]
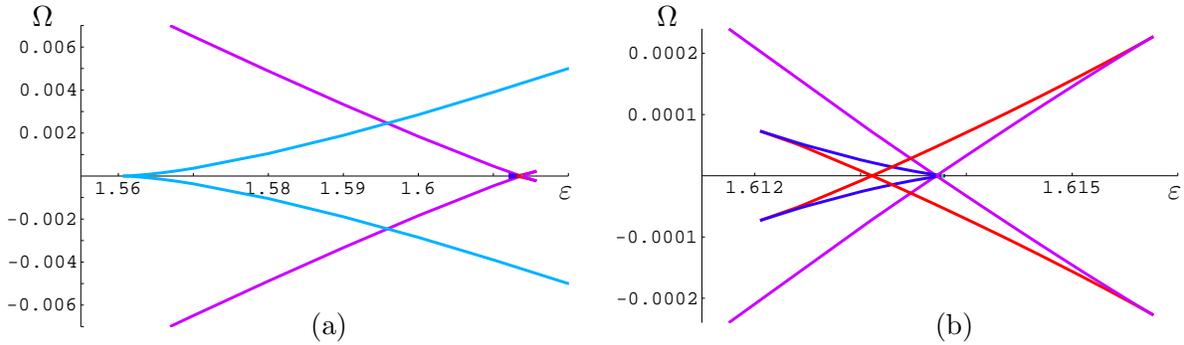

\begin{picture}(160,45)(0,0)
\put(0,1){
\psfig{file=\figdir/sndoverlap.ps,width=75mm}
}
\put(5,41){$\Omega$}
\put(75,18){$\ve$}
\put(42,0){(a)}
 \put(81,1.3){
 \psfig{file=\figdir/snddetail1.ps,width=75mm}
 }
\put(88,41){$\Omega$}
\put(156,18){$\ve$}
\put(125,0){(b)}
\end{picture}
\caption{\footnotesize
The bifurcation structure for $K = 0.8$ in the second overlap (a)
seems to be the same as in Fig. \protect\ref{fig:epsOmega}(b). However, an
enlargement (b) shows that the structure is much more complicated. 
}
\label{fig:swallowtail}
\end{figure}
\begin{figure}[t]
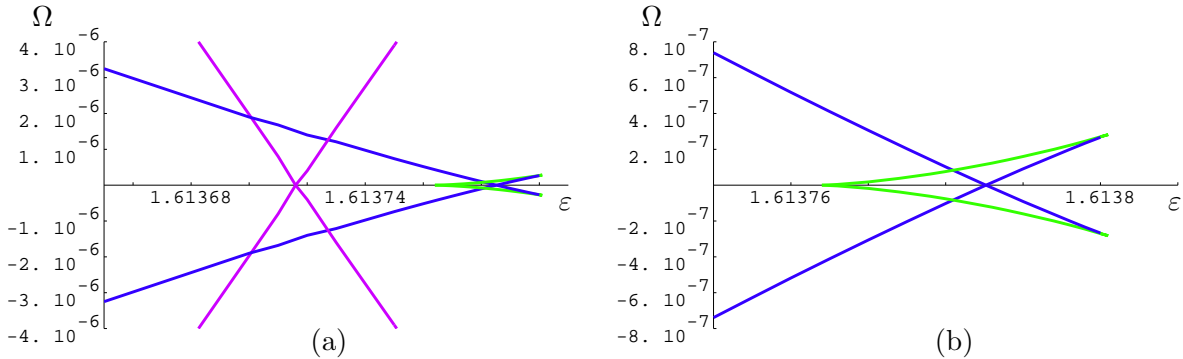

\begin{picture}(160,48)(0,0)
\put(0,0){
\psfig{file=\figdir/snddetail2.ps,width=75mm}
}
\put(5,43){$\Omega$}
\put(75,18.6){$\ve$}
\put(42,0){(a)}
\put(81,0){
\psfig{file=\figdir/snddetail3.ps,width=75mm}
}
\put(86,43){$\Omega$}
\put(156,18.6){$\ve$}
\put(125,0){(b)}
\end{picture}
\caption{\footnotesize
Details of the bifurcation structure for $K = 0.8$ in the second
overlap. The last pitchfork bifurcation on the line $\Omega = 0$ is
subcritical (b) as opposed to supercritical in
Fig.~\protect\ref{fig:epsOmega}(b).
}
\label{fig:subpf}
\end{figure}
Figure~\ref{fig:swallowtail}(a) shows a detail of the second
overlap. This picture is very similar to Fig.~\ref{fig:epsOmega}(b),
but the bifurcation diagram along $\Omega = 0$ in
Fig.~\ref{fig:bifportraits}(b) reveals a more complex structure. Let
us first discuss Fig.~\ref{fig:bifportraits}(b). The first bifurcation
(light blue dot) is the same pitchfork bifurcation as the purple dot
in Fig.~\ref{fig:bifportraits}(a): the stable circle becomes unstable,
creating two new stable circles. As $\ve$ increases two other stable
circles are born in a pair of saddle-node bifurcations (red dots). We
now have four different attractors. Note that these saddle-node
bifurcations happen at the same values of $\ve$ due to the symmetry along the
line $\Omega = 0$, as referred to earlier.

At the purple dots the two attracting circles from the pitchfork
bifurcation disappear in a pair of saddle-node bifurcations. Note that
this pair of saddle-node bifurcations is connected to the purple
pitchfork bifurcation of Fig.~\ref{fig:bifportraits}(a) via the purple
curve off $\Omega = 0$ in Fig.~\ref{fig:epsOmega}(a). We are now left
with two attractors and two repellors. These last two attractors do
not disappear in a pitchfork bifurcation as in
Fig.~\ref{fig:bifportraits}(a) (orange dot). Instead, they disappear
in a pair of saddle-node bifurcations (dark blue dots) with two
repellors that are born in a pitchfork bifurcation (green dot) for
slightly smaller $\ve$. Note that this pitchfork bifurcation is
subcritical, as opposed to the supercritical orange one in 
Fig.~\ref{fig:bifportraits}(a). 

The unfolding of these bifurcations inside the phase-locked region with $\Omega \neq 0$ is shown
in detail in Figs.~\ref{fig:swallowtail} and \ref{fig:subpf}. The
curves are colored according to the colors of the bifurcations in
Fig.~\ref{fig:bifportraits}(b). As expected, for $\Omega \neq 0$ pairs
of saddle-node bifurcations no longer happen at the same values of $\ve$. They
form two different curves that cross each other exactly at $\Omega =
0$. We already mentioned earlier that the purple curves connect all
the way left in a pitchfork bifurcation on the line $\Omega = 0$ in
the first overlap. The light blue curves start in the
pitchfork bifurcation at $\Omega = 0$ and become the outer boundary
$|\Omega| = \Omega_0(\ve, 0.8)$ once they cross the purple
curves. The red curves form swallowtails with the purple curves on the
right side and the dark blue curves on the left side; see
Fig.~\ref{fig:swallowtail}(b). Finally, the dark blue curves form 
swallowtails with red and green curves; see Fig.~\ref{fig:subpf}.

 
\section{The Structure of Bifurcations in the $(\ve, K)$-plane}
\label{sec:epsK}

\def\figoverlapb{%
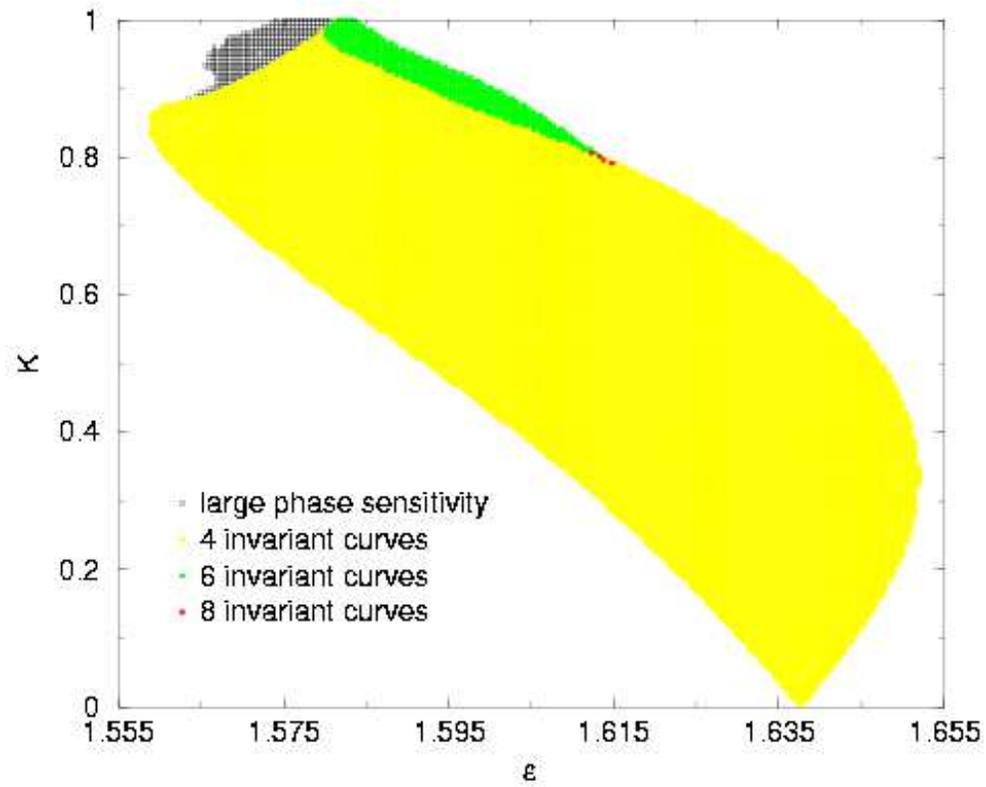
A cross-section at $\Omega = 0$ of the second overlap over a $200
\times 200$ grid. A point in the white region corresponds to two
invariant curves, one is stable and the other is unstable. The region
in which the stable curve is extremely wrinkled (large phase
sensitivity, see App.~\ref{app:sec4}) is marked as well. 
}  
\def\FIGoverlapb{\centerline{\psfig{figure=\figdir/overlap2.eps,%
width=13.0cm,angle=0}}}
\FIGo{fig:overlapb}{\figoverlapb}{\FIGoverlapb}
The bifurcation structure depends on the strength of the nonlinearity
$K$. We study the two-parameter dependence only on the cross-section
$\Omega = 0$, because the unfolding inside the phase-locked region 
with $\Omega \ne 0$ is similar to
that discussed in Sec.~\ref{sec:maintongue}. As expected, the
regions of overlap change shape with $K$. In particular, only the
regions with no more than two attractors persist for small $K$ and 
moderate $\ve$. This
is shown in Fig.~\ref{fig:overlapb} with a cross-section at $\Omega =
0$ in the $(\ve, K)$-plane of the second overlap; see
App.~\ref{app:sec4} for details on how this picture was
generated.

In the following sections we describe the bifurcations in more
detail. Section~\ref{sec:smooth} discusses the sequence of
bifurcations that happen as $K$ decreases. For small $K$ (less than 
approximately $0.8$  
for the bifurcations we have looked at),
saddle-node and pitchfork bifurcations happen via a uniform collision 
of invariant curves: at the moment of bifurcation, two (saddle-node)
or three (pitchfork) curves merge at {\it each} value of $\vt$. We
call these bifurcations {\it smooth} bifurcations. For $K$ close to 1 the
attractors may become extremely wrinkled, which gives rise to 
{\it nonsmooth} bifurcations: at the moment of bifurcation the invariant
curves now collide only in a dense set of $\vt$-values. 
The nonsmooth pitchfork and nonsmooth
saddle-node bifurcations are described in detail in
Secs.~\ref{sec:nonsmooth} and~\ref{sec:nssn}, respectively.


\subsection{Smooth bifurcations for $\Omega = 0$ in the second
            region of overlap}
\label{sec:smooth}

\begin{figure}[t]
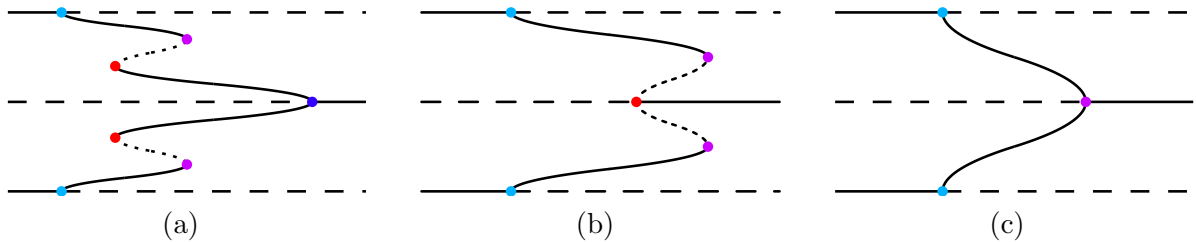

\begin{picture}(160,30)(1,0)
\put(0,5){
\psfig{file=\figdir/mingreen.ps,width=50mm}
}
\put(23,0){(a)}
\put(55,5){
\psfig{file=\figdir/minblue.ps,width=50mm}
}
\put(78,0){(b)}
\put(110,5){
\psfig{file=\figdir/minred.ps,width=50mm}
}
\put(133,0){(c)}
\end{picture}
\caption{\footnotesize
Sketches of the bifurcations along the line $\Omega = 0$ in the second
overlapping region as $K$ decreases from 0.8 to 0. The bifurcation
structure of Fig. \protect\ref{fig:bifportraits}(b) for $K = 0.8$
transforms into one like Fig. \protect\ref{fig:bifportraits}(a) by 
``absorbing'' pairs of saddle-node bifurcations in the second
pitchfork bifurcation.   
}
\label{fig:pfabsorb}
\end{figure}
Figure~\ref{fig:overlapb} indicates that at most two attractors
exist for small $K$ and moderate $\ve$. 
This means that for fixed small $K$, the
bifurcation portrait looks like Fig.~\ref{fig:bifportraits}(a). Hence,
as we decrease $K$ from $K = 0.8$ to 0, the extra pairs of saddle-node
bifurcations (see Fig.~\ref{fig:bifportraits}(b)) need to disappear somehow.
It turns out that the last pitchfork bifurcation, the green
dot in Fig.~\ref{fig:bifportraits}(b), ``absorbs'' these saddle-node
bifurcations one by one. In doing so, the pitchfork bifurcation
switches from subcritical to supercritical and vice versa (a standard 
codimension-2 bifurcation). A sketch of
this process along the line $\Omega = 0$ is shown in
Fig.~\ref{fig:pfabsorb}(a)--(c).

In the $(\Omega, \ve)$-plane the picture changes as follows. The first
swap, Fig.~\ref{fig:pfabsorb}(a), comes about as the pitchfork point
and the intersection point of the dark blue curves on $\Omega = 0$
collapse; see Fig.~\ref{fig:subpf}(b). When $K$ decreases, these
points move closer together, causing the slope of the dark blue curves
to become steeper and the ends of the swallowtail to move closer to
$\Omega = 0$. Upon collision the green curves disappear and the dark
blue curves end in a supercritical pitchfork bifurcation.

In the second swap the dark blue curves disappear in a similar way
via a collision of the pitchfork point and the intersection point of
the two red curves, making the pitchfork subcritical again. Note that,
in order for this to happen, the pitchfork point crosses the
intersection point of the two purple curves; compare
Figs.~\ref{fig:swallowtail}(b) and
\ref{fig:subpf}(a). Figures~\ref{fig:pfabsorb}(a)--(b) show why this
is a crossing and not a collision: the pair of purple saddle-node
bifurcations happens ``far out'' in state space from the pitchfork
bifurcation. Therefore, the crossing is only a crossing in this
projection on the $(\Omega, \ve)$-plane.

The third swap, Fig.~\ref{fig:pfabsorb}(c), is identical to the first,
causing the disappearance of the red curves. In this bifurcation
diagram at most two attractors coexist, which is the
desired situation for $K$ small.

\subsection{Nonsmooth pitchfork bifurcations}
\label{sec:nonsmooth}

\begin{figure}
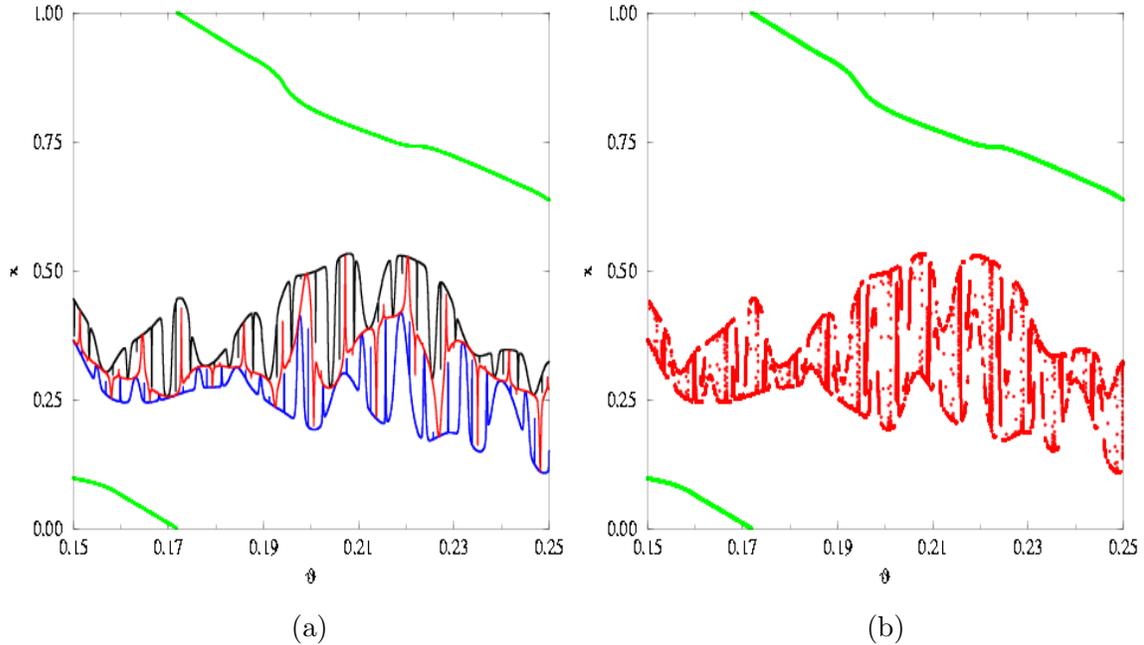

\begin{center}
\leavevmode
\psfig{figure=\figdir/before_pitch.eps,height=8.0cm,width=7.5cm,angle=0}
\psfig{figure=\figdir/after_pitch.eps,height=8.0cm,width=7.5cm,angle=0}
\begin{picture}(150,0)(0,0)
\put(38,0){(a)}
\put(114.5,0){(b)}
\end{picture}
\end{center}
\caption{\footnotesize%
Nonsmooth pitchfork bifurcation with $\Omega=0$ and $K=0.9$: (a)
before the collision ($\ve=1.56765$): attractors in blue and black,
unstable invariant curves in green and red; (b) after the collision
($\ve=1.5675$): SNA (red) and the unstable invariant curve not
taking part in the bifurcation (green).} 
\label{fig:pitch}
\end{figure}

For $K$ close to 1 the situation is more complicated, because some 
of the invariant curves are very wrinkled and the pitchfork bifurcation 
becomes nonsmooth. This bifurcation has been found by Sturman \nocite{Sturman99} [1999] 
in a similar map. Let us discuss
what happens for $\Omega = 0$, along the line $K = 0.9$ as we approach 
the pitchfork bifurcation by decreasing $\ve$, starting in the yellow region 
in Fig.~\ref{fig:overlapb}. Figure~\ref{fig:pitch}(a) shows all
invariant curves just before the bifurcation. The two stable invariant
curves (black and blue) correspond to the two outer branches of the
pitchfork, the unstable invariant curve (red) relates to the inner
branch of the pitchfork separating the two outer ones. The fourth
invariant curve (green) is also unstable, but it is ``far away'' and does
not take part in the bifurcation. As we decrease $\ve$ towards the
bifurcation point the three invariant curves (blue, black and red)
approach each other, but due to their wrinkled structures they appear to 
collide
only in a dense set of $\vt$-values instead of merging uniformly as in
a smooth bifurcation. This indicates that at the moment of bifurcation 
the attractor is an SNA. Numerical evidence suggests that this SNA persists 
and smoothes out to an invariant curve over a small $\ve$-interval; see 
for example the red attractor in Fig.~\ref{fig:pitch}(b) for $\ve$ below 
the bifurcation value. It is possible that we see the reverse of 
{\it fractalization}, a mechanism for the appearance of SNAs reported 
in \cite{NK96}. However, if we use the method of rational approximations 
for testing whether the attractor is an SNA, we get conflicting results; 
see App.~\ref{app:sec42} for more details. We remark that we get these 
conflicting numerical results only for the nonsmooth pitchfork bifurcation. 
In any case, after a 
further decrease in $\ve$ the attractor is clearly a smooth invariant curve. 

It is important to note that the nonsmooth pitchfork bifurcation is uniquely
defined as the moment of collision of three invariant curves and the 
locus of bifurcation lies on a curve in the ($\ve,K$)-plane. 
The process of fractalization is a gradual
process where the moment of transition from an invariant curve to an
SNA is not well-defined numerically. We wish to emphasize that it is, 
therefore, completely unclear whether the set of
parameter pairs $(\ve,K)$ with $\Omega = 0$ that exhibit SNAs after the 
nonsmooth pitchfork bifurcation has zero 
or finite size.  

Since the boundary of the region of bi- or multistability for $\Omega
= 0$ is given by smooth and nonsmooth pitchfork bifurcations 
there is a codimension-2 
point in the ($\ve, K$)-plane where the smooth and the nonsmooth
pitchfork bifurcation curves meet. An approximation of this
codimension-2 point is ($\ve, K$) = (1.564, 0.89). In any neighborhood of 
this point we always find all three kinds of dynamical behaviors: one 
stable invariant curve, two stable invariant curves, and one SNA; although 
the latter may only exist on the nonsmooth bifurcation curve itself.

\subsection{Nonsmooth saddle-node bifurcations}
\label{sec:nssn}

\begin{figure}
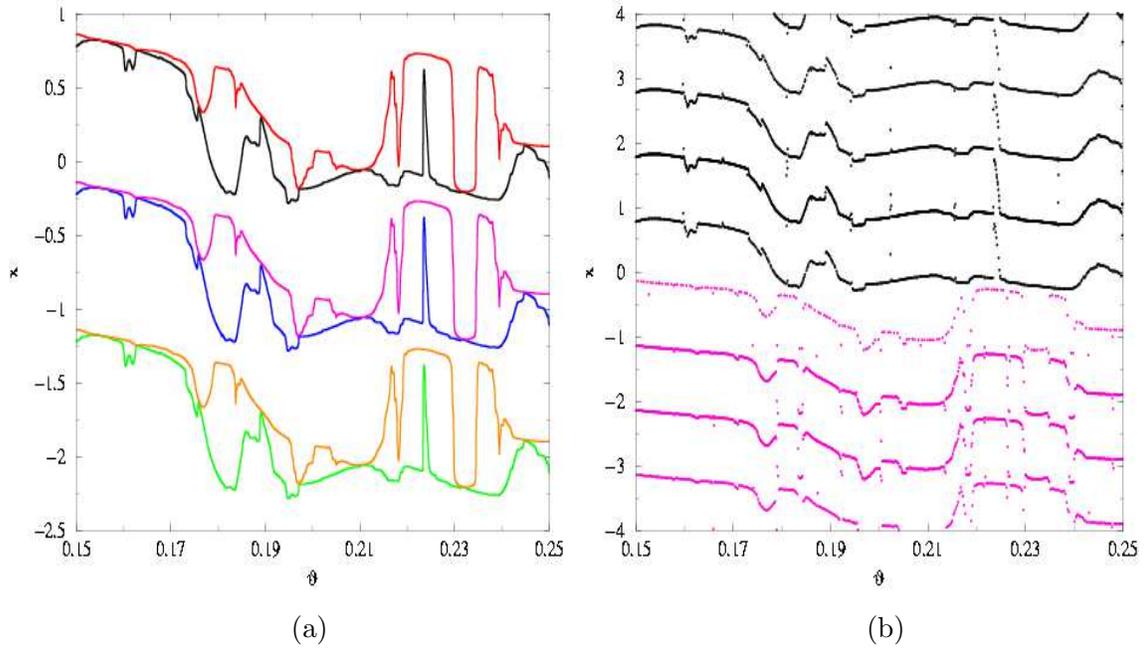

\begin{center}
\leavevmode
\psfig{figure=\figdir/before_unidir_sn.eps,height=8.0cm,width=7.5cm,angle=0}
\psfig{figure=\figdir/after_unidir_sn.eps,height=8.0cm,width=7.5cm,angle=0}
\begin{picture}(150,0)(0,0)
\put(38,0){(a)}
\put(114.5,0){(b)}
\end{picture}
\end{center}
\caption{\footnotesize%
One-sided nonsmooth saddle-node bifurcation with $\Omega=0.001$ and $\ve=2.58$: (a)
before the collision ($K=0.865$): attractors in black, blue and green,
repellors in red, purple and orange; (b) after the collision
($K=0.868$): attractor with the same initial condition as the black
attracting invariant curve in (a); repellor with the same initial condition as
the purple repelling invariant curve. The attractor moves upwards in forward
time and the repellor moves downwards in reverse time. The rotation number is
nonzero.  
}
\label{fig:osn}
\end{figure}

\begin{figure}
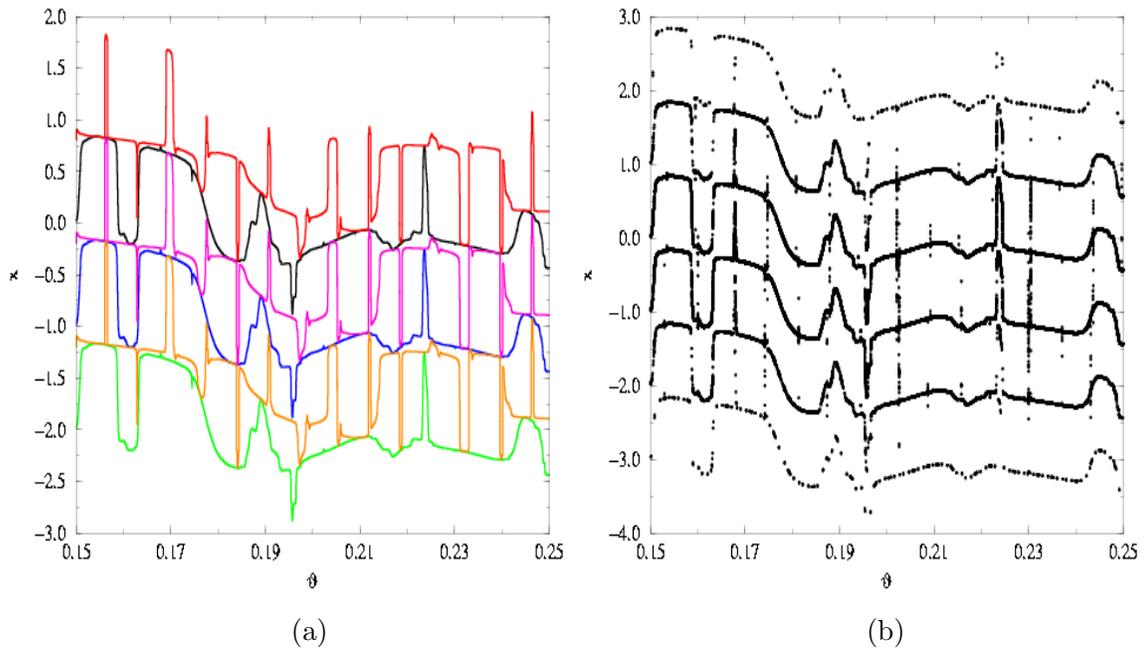

\begin{center}
\leavevmode
\psfig{figure=\figdir/before_sn.eps,height=8.0cm,width=7.5cm,angle=0}
\psfig{figure=\figdir/after_sn.eps,height=8.0cm,width=7.5cm,angle=0}
\begin{picture}(150,0)(0,0)
\put(38,0){(a)}
\put(114.5,0){(b)}
\end{picture}
\end{center}
\caption{\footnotesize%
Two-sided nonsmooth saddle-node bifurcation with $\Omega=0$ and $\ve=2.58$ 
(cf. the transition from the white to the blue region in 
Fig.~\ref{fig:overlapc}): (a)
before the collision ($K=0.927$): attractors in black, blue and green,
repellors in red, purple and orange; (b) after the collision
($K=0.928$): SNA with the same initial condition as the black
attracting invariant curve in (a). The strange repellor is not shown. 
} 
\label{fig:sn}
\end{figure}

If $K$ is small and $\ve$ moderate then the saddle-node bifurcations
observed numerically involve two invariant curves on the torus which converge
and destroy each other. At larger values of $K$, simulations suggest
that two invariant curves touch on
an orbit at the bifurcation point, so points of intersection of these sets are
dense on the curves. (Strictly speaking, the invariant sets are no longer 
continuous at the bifurcation point, but we will continue to refer to 
them as curves.) For quasiperiodically forced circle maps we 
can distinguish two types of these nonsmooth saddle-node bifurcations: 
one-sided and two-sided. 
In the one-sided nonsmooth saddle-node bifurcation, the collisions  
occur between pairs of invariant curves on the cylinder. An example is
shown in Fig.~\ref{fig:osn}(a) where the invariant curves and some of their
translates by one in the $x-$direction are computed close to the
bifurcation point. In the two-sided case each stable invariant
curve on the cylinder touches both the unstable invariant curve 
immediately above it and the unstable invariant curve immediately below it.  
On the torus this implies that at the bifurcation point the 
attractor is everywhere discontinuous. An example of a two-sided nonsmooth 
saddle-node bifurcation is shown in Fig.~\ref{fig:sn}(a). 
These two-sided nonsmooth 
saddle-node bifurcations are described in \cite{FKP95}, where it is shown 
that after the bifurcation (with $\Omega =0$ and $K$ fixed)
the map has an SNA with unbounded motion in the $x-$direction 
(Fig.~\ref{fig:sn}(b)) despite
the fact that the rotation number remains zero due to the symmetry of Eqs.~(\ref{eq:mapx})--(\ref{eq:maptheta})
when $\Omega =0$. This implies that the diffusion in the 
$x-$direction is extremely slow; see \cite{FKP95,SFGP99} for further details. 
In general, the two-sided nonsmooth saddle-node bifurcation is
of codimension two, but it occurs as a codimension-1 phenomenon due to the 
symmetry if $\Omega =0$. Figure~\ref{fig:overlapc} shows the range of dynamics 
observed in the third region of overlap in the plane $\Omega = 0$. Unbounded 
SNAs are observed in the blue regions and two-sided nonsmooth saddle-node 
bifurcations occur on the boundary between the white and blue regions. 

The unbounded SNA of Fig.~\ref{fig:sn}(b) must contain orbits which are unbounded
above and orbits which are unbounded below \cite{SFGP99}. 
This bidirectional diffusive motion of the unbounded
SNA with $\Omega =0$ becomes effectively unidirectional if $|\Omega |$
is very small, leading to a nonzero rotation number \cite{SFGP99}.
This suggests that these unbounded SNAs lie on the boundary of the 
phase-locked region. Fig.~\ref{fig:tongue} shows this boundary 
in $(\ve ,K,\Omega)$-
space. It is clear that the height (i.e. the width in $\Omega$)
of the boundary is very small, if not zero, in regions of the $(\ve,K)$
plane with $\Omega =0$ which have unbounded SNAs (compare the low plateau
on the left of Fig.~\ref{fig:tongue}  with the region of unbounded SNAs 
of Fig.~\ref{fig:overlapc}). As
the height of the boundary becomes non-negligible we observe that the
saddle-node bifurcation on the boundary has become one-sided (see
Fig.~\ref{fig:osn}), and we 
believe that it remains one-sided and nonsmooth throughout the red 
areas of Fig.~\ref{fig:tongue} with non-negligible height. These red 
regions of 
the boundary correspond to saddle-node bifurcations with negative
nontrivial Lyapunov exponents, and appear to preceed the creation of
SNAs with nonzero rotation numbers outside the phase-locked region 
\cite{DGO89,Glendinning98}.

If the phase-locked region really has zero height on the plateau, then
it is wrong to refer to saddle-node bifurcations on the interior of the 
plateau: these points would correspond to a transition from an SNA
with negative rotation number to an SNA with positive rotation number through 
an SNA with zero rotation number as $\Omega$ increases through zero. In this
full three-parameter unfolding the two-sided nonsmooth saddle-node bifurcations are
of codimension two, occurring on curves bounding the plateau and separating
parts of the boundary of the phase-locked region with unbounded SNAs from
parts with one-sided nonsmooth saddle-node bifurcations.

\def\figoverlapc{%
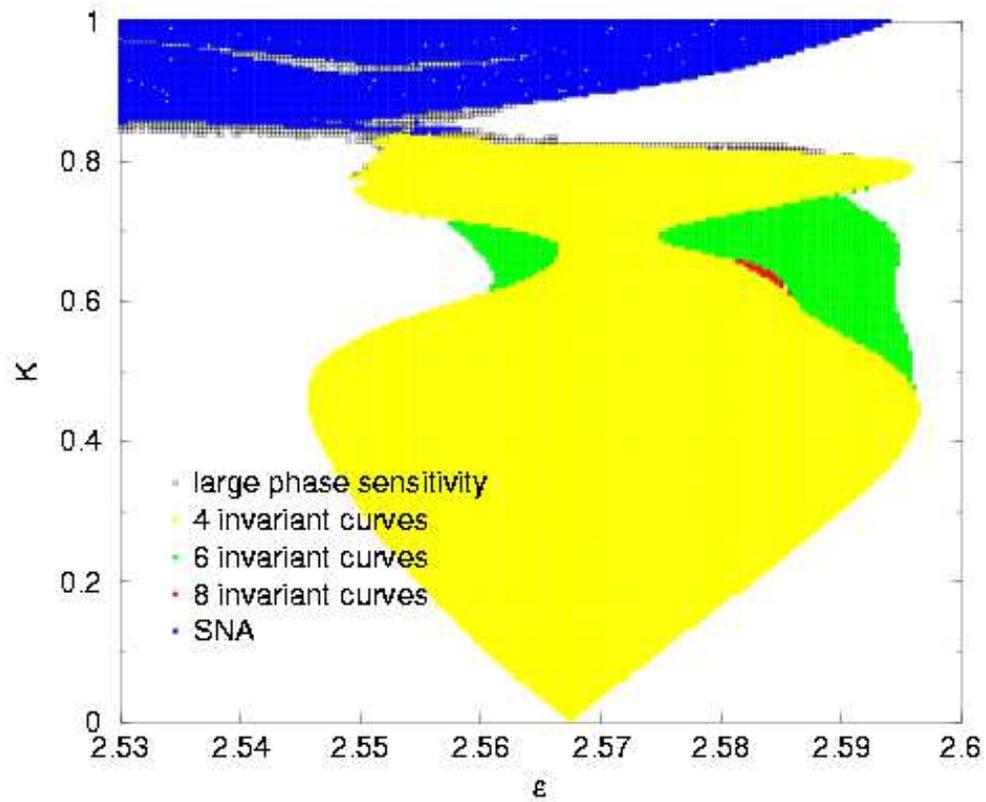
Section $\Omega = 0$ of the third overlap region. In the region of
large phase sensitivity the attractor is very wrinkled or even an
SNA.} 
\def\FIGoverlapc{%
\centerline{\psfig{figure=\figdir/overlap3.eps,width=13.0cm,angle=0}}}
\FIGo{fig:overlapc}{\figoverlapc}{\FIGoverlapc}
\def\figtongue{%
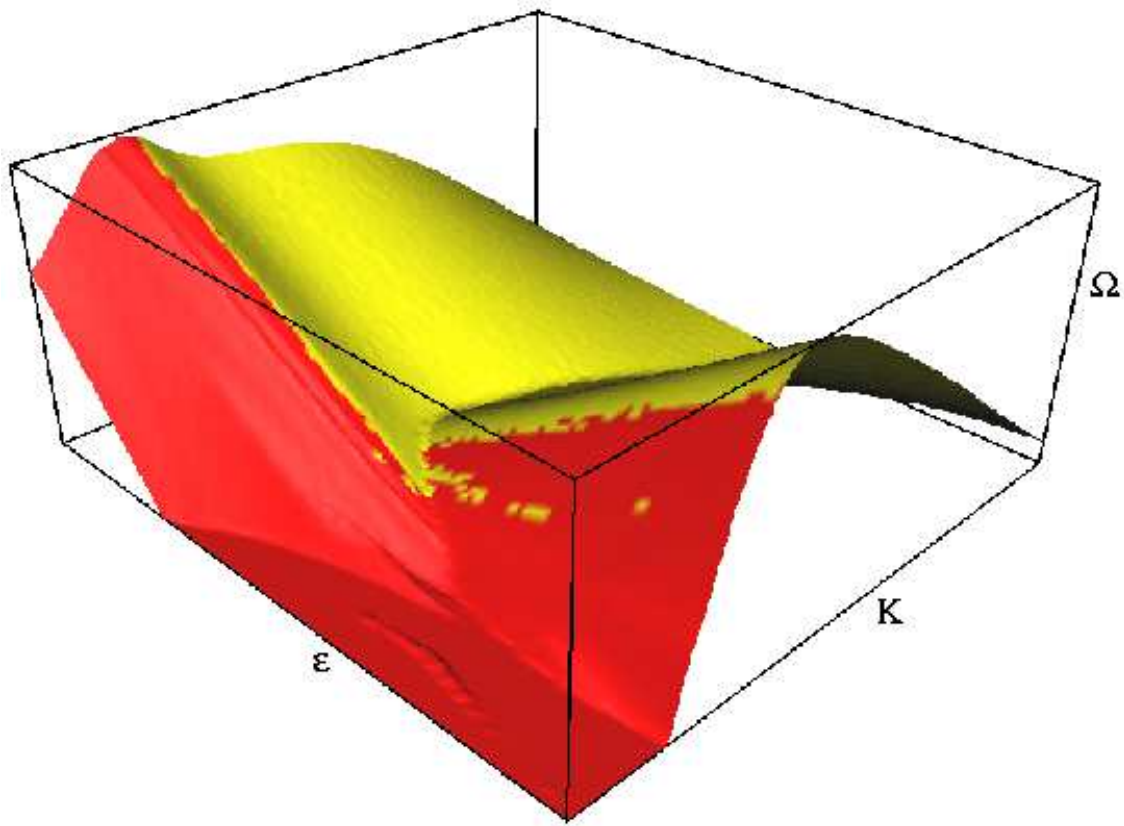
A small part of the boundary $|\Omega| = \Omega_0(\varepsilon, K)$ of 
the phase-locked region with zero rotation number; $\varepsilon \in
[2.52, 2.62]$ runs from right to left, $K \in [0, 1]$ also from right
to left, and $|\Omega| \in [0,0.004]$  from bottom to top; compare
\fig~\protect\ref{fig:overlapc}. A grid of $80 \times 160$ points in
the $(\ve, K)$-plane is taken. Red regions with negative
Lyapunov exponent (using $\lambda < -0.005$ as criterion) correspond
to transitions to SNAs (the yellow spots are due to the finite
threshold).
}
\def\FIGtongue{%
\centerline{\psfig{figure=\figdir/tongue.eps,width=15.0cm,angle=0}
}}
\FIGo{fig:tongue}{\figtongue}{\FIGtongue}

\section{Nonsmooth bifurcation points of codimension two}
\label{sec:nscodim}

%
\def\figoverlapcmag{%
(a) Magnification of \fig~\protect\ref{fig:overlapc}. The white region
corresponds to pairs of invariant curves, regardless of their phase
sensitivity. (b) Magnification of (a).} 
\def\FIGoverlapcmag{\centerline{\psfig{figure=\figdir/mag.eps,width=16.0cm,angle=0}
}}
\FIGo{fig:overlapcmag}{\figoverlapcmag}{\FIGoverlapcmag}
In the full parameter space, and also already in the $(\ve, K)$-plane
with $\Omega = 0$, we expect codimension-2 bifurcations. We have
already seen some of these; for example, the transition from yellow to
red in Figs.~\ref{fig:maintongue} and~\ref{fig:tongue} is the
codimension-2 bifurcation curve marking the transition from a smooth
saddle-node bifurcation to a nonsmooth one, respectively~\cite{KNPS00}. Another
example is the codimension-2 point, mentioned in
Sec.~\ref{sec:nonsmooth}, where the smooth and nonsmooth pitchfork
bifurcation curves meet. Note that there is no {\em curve} of
codimension-2 points in this case, since the pitchfork bifurcation is
restricted to the plane $\Omega = 0$. As discussed in the previous 
section, there is also the two-sided nonsmooth saddle-node bifurcation curve.

In this section we want to draw attention to an interesting
codimension-2 point in the plane $\Omega = 0$. This point
can be seen in Fig.~\ref{fig:overlapc} and in the enlargements
Figs.~\ref{fig:overlapcmag}(a)--(b) as the point
where the region with unbounded SNAs (blue) and the bistable region
(yellow) touch. It can be characterized as the moment where two
nonsmooth (supercritical) pitchfork bifurcations happen
simultaneously, i.e.\ a region of overlap is pulled apart; compare
also the sketch in Fig.~7 of~\cite{GFPS99}.

\subsection{Smooth analog of the nonsmooth codimension-2 bifurcation
point} 
\label{sec:analog}

\begin{figure}[t]
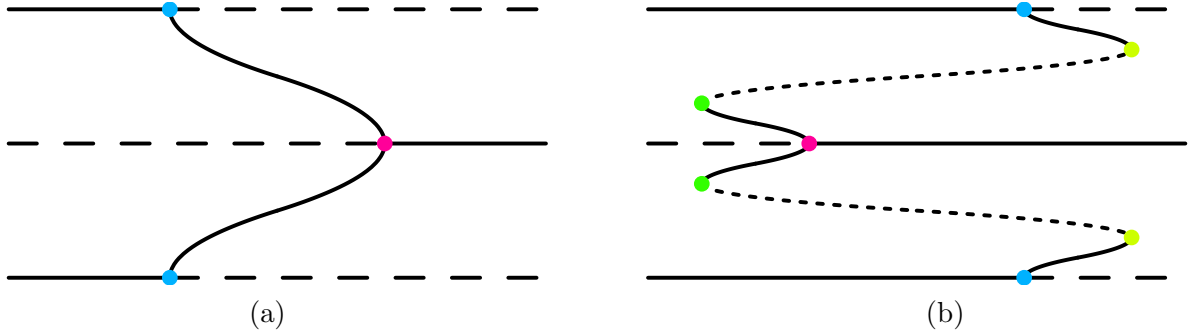

\begin{picture}(160,45)(0,0)
\put(0,5){
\psfig{file=\figdir/analog1.ps,width=75mm}
 }
\put(35,0){(a)}
\put(85,5){
\psfig{file=\figdir/analog2.ps,width=75mm}
}
\put(125,0){(b)}
\end{picture}
\caption{\footnotesize
Two different bifurcation diagrams along the line $\Omega = 0$ that
both involve two (smooth) supercritical pitchfork bifurcations. Shown are
invariant curves (represented by one point) versus $\ve$ with $K$
fixed. If the pitchfork bifurcation creating two attractors does not
happen first, we necessarily need extra curves of saddle-node
bifurcations~(b).
}
\label{fig:analog}
\end{figure}
Let us first discuss the smooth analog of this nonsmooth codimension-2
bifurcation point. Suppose for $\Omega = 0$ and some $K < 1$ fixed the
bifurcation diagram involves only two supercritical pitchfork
bifurcations that occur in the order as shown in
Fig.~\ref{fig:analog}(a). Now assume that as we increase $K$, this
order is switched before we reach $K = 1$, without changing the type
of pitchfork bifurcation from supercritical to subcritical. As shown
in Fig.~\ref{fig:analog}(b), this means that we necessarily need extra
curves of saddle-node bifurcations. 

\begin{figure}[t]
\begin{center}
\begin{picture}(80,85)(0,5)
\put(0,5){
\psfig{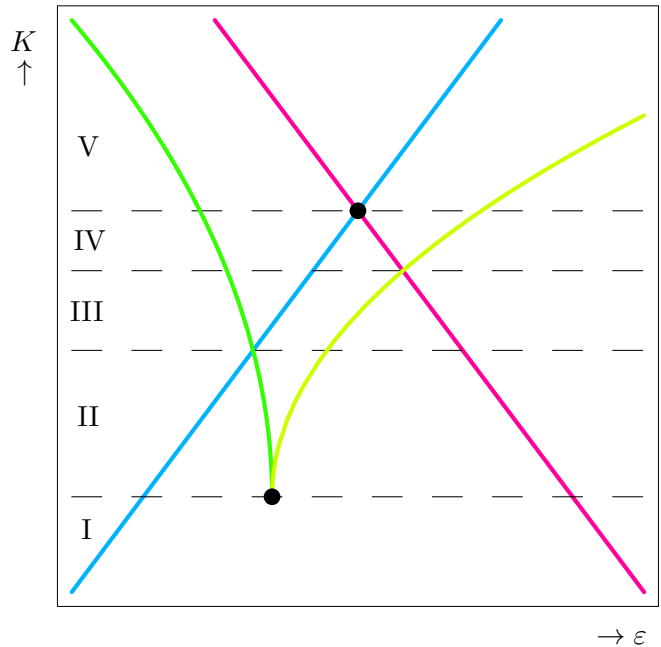}
}
\put(-5,79){$K$}
\put(-4,75){$\uparrow$}
\put(73,0){$\rightarrow \ve$}
\put(4.5,14){I}
\put(4,29){II}
\put(3,43){III}
\put(3.5,52.5){IV}
\put(4,65){V}
\end{picture}
\end{center}
\caption{\footnotesize
Unfolding of the smooth analog of the nonsmooth codimension-2
bifurcation point. Green and light-green curves are saddle-node bifurcations, 
light-blue and magenta curves correspond to pitchfork bifurcations. 
}
\label{fig:smooth-unfold}
\end{figure}
The smooth analog of the nonsmooth
codimension-2 bifurcation point is the point where the two pitchfork
bifurcations happen at the same parameter values.
In the $(\ve, K)$-plane the complete bifurcation diagram should look
like Fig.~\ref{fig:smooth-unfold}. The two supercritical pitchfork
bifurcation curves are colored light-blue and magenta. The green and
light-green curves are saddle-node bifurcation curves; compare also
the colors in Figs.~\ref{fig:analog}(a)--(b). Above the green and
light-green curves, but below the light-blue and magenta curves there
are four attractors. If we cross either the light-blue or the magenta
curve from this region there are three attractors. Above the
light-blue and magenta curves there are two attractors. 
\begin{figure}[t]
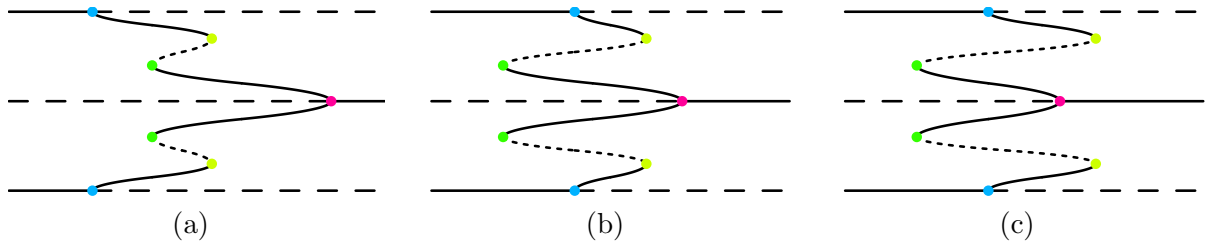

\begin{picture}(160,30)(1,0)
\put(0,5){
\psfig{file=\figdir/regionII.ps,width=50mm}
}
\put(23,0){(a)}
\put(55,5){
\psfig{file=\figdir/regionIII.ps,width=50mm}
}
\put(78,0){(b)}
\put(110,5){
\psfig{file=\figdir/regionIV.ps,width=50mm}
}
\put(133,0){(c)}
\end{picture}
\caption{\footnotesize
Qualitative bifurcation diagrams for $K$-values in the $K$-intervals
II (a), III (b), and IV (c) in Fig.~\protect\ref{fig:smooth-unfold}. 
The equivalent diagrams for the $K$-intervals I and V are shown in 
Fig.~\ref{fig:analog}. 
}
\label{fig:regII-IV}
\end{figure}
We distinguish five qualitatively different $K$-intervals numbered
I--V in Fig.~\ref{fig:smooth-unfold}. Interval I corresponds to
Fig.~\ref{fig:analog}(a) and V to Fig.~\ref{fig:analog}(b). The
qualitative behavior in the intervals II--IV is given in
Figs.~\ref{fig:regII-IV}(a)--(c), respectively. The codimension-2
point that we are discussing here is the intersection point of the
light-blue and magenta pitchfork bifurcation curves. The intersections
of the green and light-blue curves, and the light-green and magenta
curves are only intersections in this projection onto the $(\ve,
K)$-plane as can be seen in Fig.~\ref{fig:regII-IV}, where the
transition from Fig.~\ref{fig:regII-IV}(a) to 
(b) marks the crossing of green and light-blue,
and the transition from Fig.~\ref{fig:regII-IV}(b) to 
(c) represents the crossing of light-green and
magenta. 

\subsection{A nonsmooth bifurcation point of codimension two} 
\label{sec:codim2}

\begin{figure}[t]
\begin{center}
\begin{picture}(40,45)(0,5)
\put(0,5){
\psfig{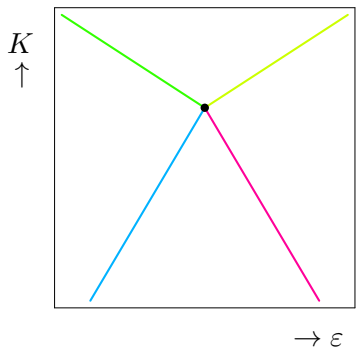}
}
\put(-5,39){$K$}
\put(-4,35){$\uparrow$}
\put(33,0){$\rightarrow \ve$}
\end{picture}
\end{center}
\caption{\footnotesize
Unfolding of the nonsmooth codimension-2 bifurcation point. The green and 
light-green curves correspond to nonsmooth saddle-node bifurcations, while the 
light-blue and the magenta curves represent nonsmooth pitchfork bifurcations. 
}
\label{fig:nonsmooth-unfold}
\end{figure}
The nonsmooth version of Fig.~\ref{fig:smooth-unfold} looks
surprisingly simple in contrast; see
Fig.~\ref{fig:nonsmooth-unfold}. The coloring of the bifurcation
curves is as in Fig.~\ref{fig:smooth-unfold} with the restriction that
all curves represent nonsmooth
bifurcations. The upper pair of bifurcation curves corresponds to two-sided
nonsmooth saddle-node bifurcations and the lower pair to nonsmooth 
pitchfork bifurcations. 

\begin{figure}
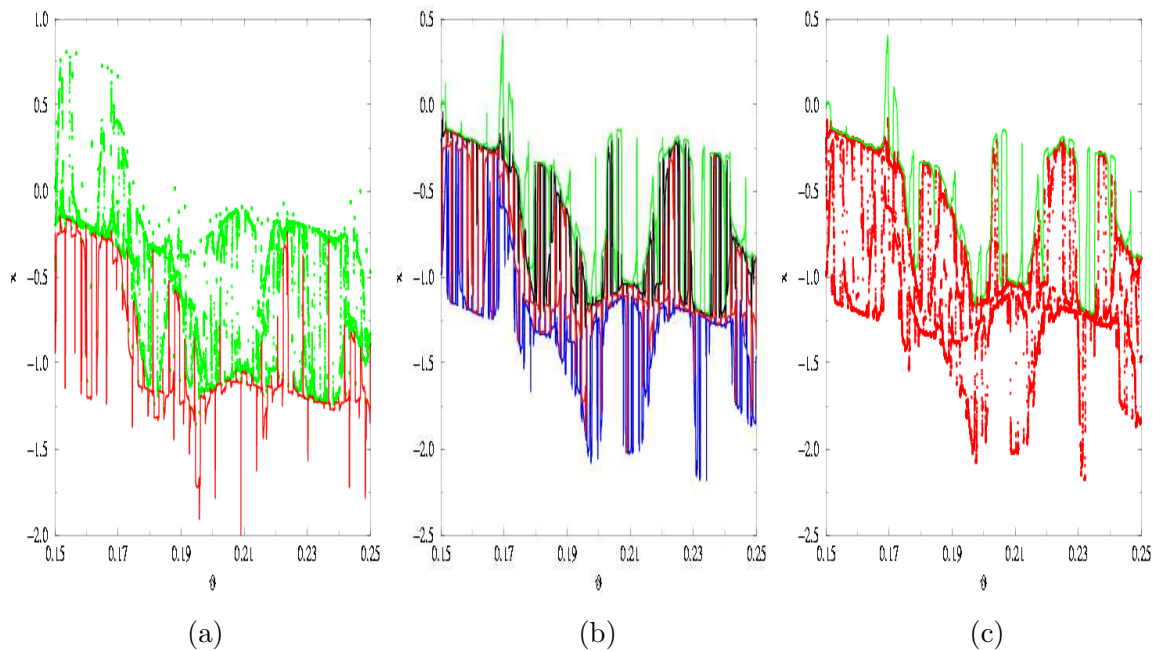

\begin{center}
\leavevmode
\psfig{figure=\figdir/left_codim2.eps,height=8.0cm,width=5.0cm,angle=0}
\psfig{figure=\figdir/before_codim2.eps,height=8.0cm,width=5.0cm,angle=0}
\psfig{figure=\figdir/right_codim2.eps,height=8.0cm,width=5.0cm,angle=0}
\begin{picture}(150,0)(0,0)
\put(23,0){(a)}
\put(75,0){(b)}
\put(127,0){(c)}
\end{picture}
\end{center}
\caption{\footnotesize%
Nonsmooth pitchfork bifurcations in the neighborhood of the
codimension-2 point $K=0.83811$, $\Omega=0$ and $\ve=2.55483$: (a)
to the left of the bistable region ($K=0.838$, $\ve=2.5542$); (b) within the
bistable region close to the higher codimension point ($K=0.838$,
$\ve=2.55483$); (c) to the right of the bistable region ($K=0.838$,
$\ve=2.5551$).
} 
\label{fig:codim2}
\end{figure}
We now have four regions with distinctive dynamics. To the left and to the
right of the two pitchfork bifurcation curves there is one attractor. 
In between the two curves of pitchfork bifurcations there are two
attractors. This is illustrated for a particular choice of the
parameters in Fig.~\ref{fig:codim2}(b). The two attractors (black and
blue) are two invariant curves separated by unstable invariant curves
(green and red). In contrast to the previously discussed case of the
nonsmooth pitchfork bifurcation, both unstable invariant curves are now
close to the attracting invariant curves, because we are near 
the codimension-2 bifurcation point. 
The two different pitchfork bifurcations are 
illustrated in Figs.~\ref{fig:codim2}(a) and (c). 
If we decrease the forcing amplitude $\ve$, the black, blue and
green invariant curves disappear in a nonsmooth pitchfork bifurcation
to form the green attractor coexisting with the red unstable
invariant curve, as shown in Fig.~\ref{fig:codim2}(a). This
corresponds to a crossing of the light-blue nonsmooth pitchfork
bifurcation curve in Fig.~\ref{fig:nonsmooth-unfold}. We cross the
magenta nonsmooth pitchfork bifurcation curve in this figure by
increasing $\ve$. In this case the red, black and blue invariant
curves disappear and form the red attractor coexisting with the
green unstable invariant curve as in Fig.~\ref{fig:codim2}(c). The
fourth region between the nonsmooth saddle-node bifurcations 
is characterized by the existence of unbounded SNAs.

Finally, we remark that the SNA region seems to have a fractal-like 
structure in
the neighborhood of the codimension-2 point; compare
Figs.~\ref{fig:overlapcmag}(a)--(b). This would
imply that the boundary of the phase-locked region is fractal!

 
\section{Tongues with Nonzero Rotation Numbers}
\label{sec:neighbourhood}

\def\fignonzero{%
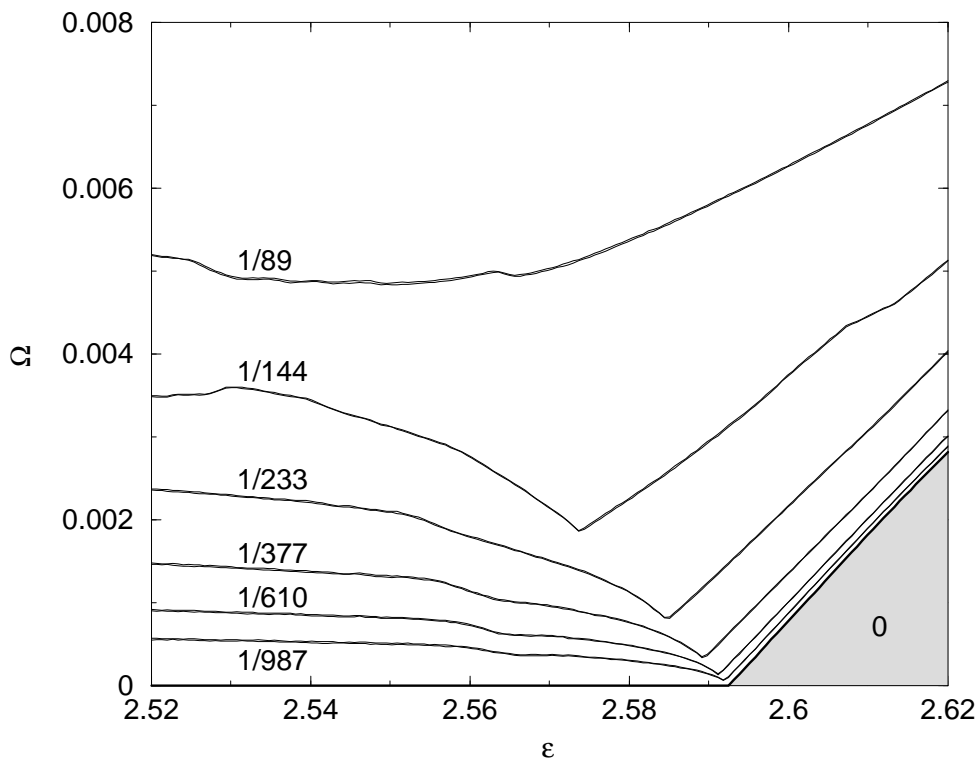
Section $K = 0.99$ of the regions with rotation numbers $\rho =
1/F_k$, $k$ being $11, \ldots, 16$, close to the (grey) region with
$\rho = 0$; compare \figs~\protect\ref{fig:overlapc}--\protect\ref{fig:tongue}. 
}  
\def\FIGnonzero{%
\centerline{\psfig{figure=\figdir/nonzero.eps,width=13.0cm,angle=0}
}}
\FIGo{fig:nonzero}{\fignonzero}{\FIGnonzero}
We studied the structure of the phase-locked region with zero rotation
number in great detail. In this section we briefly discuss the
geometry of the tongues with nonzero rotation numbers that are close to the main tongue. More precisely, we determine the boundaries of the
tongues with rotation numbers $\rho = 1/F_k$, with $F_k$ the $k$th Fibonacci
number $F_k = F_{k-1} + F_{k-2}$ and $F_1 = F_2 = 1$, using the same numerical
procedure as before; see App.~\ref{app:sec22}. Several tongues for
$K = 0.99$ are shown in Fig.~\ref{fig:nonzero}. The fluctuations of
the widths of the tongues with $\rho > 0$ are due to numerical errors,
which are of the same magnitude as the widths themselves. The thick
borderline $\Omega_0(\varepsilon, K)$ in this parameter regime
corresponds to transitions to SNAs or bifurcations of SNAs of the type 
discussed in Sec.~\ref{sec:nssn}. 

\def\figscaling{%
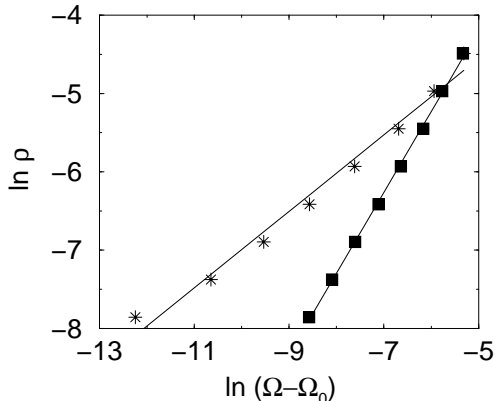
The distance $\Omega_\rho - \Omega_0$ of the tongue with rotation
number $\rho = 1/F_k$ to the main tongue, where $k$ ranges from 11
to 18; compare also \fig~\protect\ref{fig:nonzero}. The data can be
approximated by straight lines in a $\ln-\ln$ plot with slopes $1.032
\pm 0.013$ for $\varepsilon = 2.55$ (squares) and $0.488 \pm 0.025$
for $\varepsilon = 2.61$ (stars). The nonlinearity $K$ is in both
cases equal to $0.99$.
} 
\def\FIGscaling{%
\centerline{\psfig{figure=\figdir/scaling.eps,width=6.5cm,angle=0}
}}
\FIGo{fig:scaling}{\figscaling}{\FIGscaling}
How these tongues approach the boundary $\Omega_0(\varepsilon, K)$
for fixed $(\varepsilon, K)$ as $k$ goes to infinity depends in
general on $\varepsilon$ and $K$; compare $\varepsilon = 2.55$ and
$\varepsilon = 2.61$ in \fig~\ref{fig:nonzero}. This behavior can be
quantified with a scaling law for the distance $\Omega_\rho -
\Omega_0$ between the main tongue and the tongues with nonzero
rotation numbers $\rho$. We used several parameter pairs
$(\varepsilon, K)$ in \fig~\ref{fig:scaling} and our numerical
calculations strongly suggest the scaling law $\rho \sim \Omega_\rho$
for $\ve$ and $K$ such that $\Omega_0 = 0$, and $\rho \sim
\sqrt{\Omega_\rho - \Omega_0}$ otherwise. This agrees with the linear
scaling in the high-$\varepsilon$ limit conjectured by Ding et
al.~\nocite{DGO89} [1989], the square-root scaling in the \arn circle map for
$\varepsilon = 0$ and $K > 0$ (see e.g.~\cite{MacKayTresser84}) and the trivial linear
scaling of the pure rotation for $\varepsilon = K = 0$. 
The square root scaling is associated with the one-sided 
saddle-node bifurcation, and the linear scaling is related to the change 
in the rotation number of the SNAs as in Sec.~\ref{sec:nssn}. 
The change between these scalings is associated with a two-sided nonsmooth 
saddle-node bifurcation point of codimension two at which two one-sided
nonsmooth saddle-node  bifurcation curves meet (presumably the nonsmooth
analog of a cusp bifurcation point). 
A detailed study of this curve of codimension-2 points could be of interest.

 
\section{Summary}
\label{sec:disc}

We have studied the structure of the phase-locked regions in the 
quasiperiodically forced circle map. In particular, we have found 
regions of multistability where several attractors coexist. These 
regions of multistability appear due to the emergence of additional 
pairs of invariant curves as a result of saddle-node or pitchfork 
bifurcations under the variation of the forcing amplitude. As a result, 
these regions look like overlaps of phase-locked regions with the same 
rotation number. 

Opening and closing of these pockets of multistability are due to
saddle-node and pitchfork bifurcations of invariant curves. These
bifurcations can be either smooth or nonsmooth depending on the
strength of nonlinearity and the forcing amplitude. This is organized
by the type of interaction between the stable and unstable invariant
curves. In the smooth case these curves approach each other uniformly
and then touch uniformly in each value of $\vt$, so that the
bifurcation looks like a simple merging of the invariant curves, analogous 
to the unforced case. Nonsmooth saddle-node bifurcations appear due to a 
wrinkled structure of the participating invariant curves, which
collide at the bifurcation only in a dense set of $\vt$-values. The
result of this bifurcation is the emergence of a strange nonchaotic
attractor. Similar to the nonsmooth saddle-node bifurcation we find a
nonsmooth pitchfork bifurcation, but the details of this  
bifurcation are still unclear. Both for the saddle-node and the pitchfork
bifurcation there are codimension-2 points in parameter space marking the
transition from a smooth to a nonsmooth bifurcation. 
The exact determination of the codimension-2 pitchfork bifurcation point and the 
self--similarity properties in its neighborhood should be possible by
using renormalization group techniques. 

We have also investigated a nonsmooth codimension-2 bifurcation involving
the merging of two nonsmooth pitchfork
bifurcations and two nonsmooth saddle-node bifurcations which leads 
to regions of unbounded SNAs. This has no straightforward analog in smooth 
bifurcations. 

The positions of phase-locked regions in the 
neighborhood of the region with zero rotation number were also described. 
Fixing the 
nonlinearity $K$ and the forcing $\varepsilon$, we found that the 
rotation number $\rho$ of these regions scales linearly with
$\Omega_{\rho}$ whenever the width of the main tongue appears to be zero, 
and as $\sqrt{\Omega_{\rho} - \Omega_0}$ otherwise.

 
\vspace{1cm}

{\Large
\noindent
{\bf Acknowledgements}}

\vspace{3mm}

\noindent
We thank for the hospitality of the Max Planck Institute for Physics
of Complex Systems in Dresden, where this collaboration started during the
Workshop and Seminar ``Beyond Quasiperiodicity: Structures and Complex
Dynamics'' (January 1999), and where the work was completed in March 2000. 
H.O. was supported by an AFOSR/DDRE MURI grant
AFS-5X-F496209610471 while she was employed at Caltech in Pasadena,
USA. J.W. acknowledges financial support by the EU through a TMR
Network on the dynamics of spatially extended systems under contract
number FMRXCT960010. P.G. is grateful for a travel grant from the British
Council. U.F. acknowledges financial support by the
Deutsche Forschungsgemeinschaft (Heisenberg-Program and Sfb 555).


\bibliographystyle{ijbc}
{\small
\bibliography{}

\begin{thebibliography}{}

\bibitem[Arnol$'$d, 1965]{Arnold65}
Arnol$'$d, V.~I. [1965] {``Small denominators {I},''} {\em Trans. Amer. Math.
  Soc. 2nd. series} {\bf 46}, 213--284.

\bibitem[Arnol$'$d, 1983]{Arnold83}
Arnol$'$d, V.~I. [1983] {``Remarks on the perturbation-theory for problems of
  mathieu type,''} {\em Russian Mathematical Surveys} {\bf 38}, 215--233.

\bibitem[Bohr et~al., 1984]{BBJ84}
Bohr, T., Bak, P.  \& Jensen, M.~H. [1984] {``Transition to chaos by
  interaction of resonances in dissipative systems. {II}. {J}osephson
  junctions, charge-density waves, and standard maps,''} {\em Phys. Rev. A}
  {\bf 30}, 1970--1981.

\bibitem[Bohr et~al., 1985]{BBJ85}
Bohr, T., Bak, P.  \& Jensen, M.~H. [1985] {``Mode-locking and the transition
  to chaos in dissipative systems,''} {\em Physica Scripta} {\bf T9}, 50--58.

\bibitem[Ding et~al., 1989]{DGO89}
Ding, M., Grebogi, C.  \& Ott, E. [1989] {``Evolution of attractors in
  quasiperiodically forced systems: From quasiperiodic to strange nonchaotic to
  chaotic,''} {\em Phys. Rev. A} {\bf 39}(5), 2593--2598.

\bibitem[Feudel et~al., 1997]{FGO97}
Feudel, U., Grebogi, C.  \& Ott, E. [1997] {``Phase-locking in
  quasiperiodically forced systems,''} {\em Physics Reports} {\bf 290}, 11--25.

\bibitem[Feudel et~al., 1995]{FKP95}
Feudel, U., Kurths, J.  \& Pikovsky, A. [1995] {``Strange non-chaotic attractor
  in a quasiperiodically forced circle map,''} {\em Physica} {\bf 88D},
  176--186.

\bibitem[Glendinning, 1998]{Glendinning98}
Glendinning, P. [1998] {``Intermittency and strange nonchaotic attractors in
  quasiperiodically forced circle maps,''} {\em Phys.~Lett. A} {\bf 244},
  545--550.

\bibitem[Glendinning et~al., 2000]{GFPS99}
Glendinning, P., Feudel, U., Pikovsky, A.  \& Stark, J. [2000] {``The structure
  of mode-locked regions in quasi-periodically forced circle maps,''} (to
  appear in Physica D).

\bibitem[Glendinning \& Wiersig, 1999]{GW99}
Glendinning, P. \& Wiersig, J. [1999] {``Fine structure of mode-locked regions
  of the quasi-periodically forced circle map,''} {\em Phys.~Lett. A} {\bf
  257}, 65--69.

\bibitem[Grebogi et~al., 1984]{GOPY84}
Grebogi, C., Ott, E., Pelikan, S.  \& Yorke, J.~A. [1984] {``Strange attractors
  that are not chaotic,''} {\em Physica} {\bf 13D}, 261--268.

\bibitem[Hall, 1984]{Hall84}
Hall, G.~R. [1984] {``Resonance zones in two-parameter families of circle
  homeomorphisms,''} {\em SIAM J. Math. Anal.} {\bf 15}(6), 1075--1081.

\bibitem[Jensen et~al., 1983]{JBB83}
Jensen, M.~H., Bak, P.  \& Bohr, T. [1983] {``Complete devil's staircase,
  fractal dimension, and universality of mode-locking structure in the circle
  map,''} {\em Phys. Rev. Lett.} {\bf 50}, 1637--1639.

\bibitem[Jensen et~al., 1984]{JBB84}
Jensen, M.~H., Bak, P.  \& Bohr, T. [1984] {``Transition to chaos by
  interaction of resonances in dissipative systems. {I}. circle maps,''} {\em
  Phys. Rev. A} {\bf 30}, 1960--1969.

\bibitem[Kuznetsov et~al., 2000]{KNPS00}
Kuznetsov, S., Neumann, E., Pikovsky, A.  \& Sataev, I. [2000] {``Critical
  point of tori-collision in quasiperiodically forced systems,''} preprint,
  University of Potsdam (submitted to Phys. Rev. E).

\bibitem[MacKay \& Tresser, 1984]{MacKayTresser84}
MacKay, R.~S. \& Tresser, C. [1984] {``Transition to chaos for two-frequency
  systems,''} {\em J. Physique Lett.} {\bf 45}, L741--L746.

\bibitem[McGehee \& Peckham, 1996]{McGP96}
McGehee, R.~P. \& Peckham, B.~B. [1996] {``Arnold flames and resonance surface
  folds,''} {\em Int. J. Bif. and Chaos} {\bf 6}, 315--336.

\bibitem[Nishikawa \& Kaneko, 1996]{NK96}
Nishikawa, T. \& Kaneko, K. [1996] {``Fractalization of torus revisited as a
  strange nonchaotic attractor,''} {\em Phys. Rev. E} {\bf 54}(6), 6114--6124.

\bibitem[Osinga et~al., 2000]{OWGFwww00}
Osinga, H., Wiersig, J., Glendinning, P.  \& Feudel, U. [2000]
  {``Multistability and nonsmooth bifurcations in the quasiperiodically forced
  circle map,''} {\em
  http://www.mpipks-dresden.mpg.de/eprint/jwiersig/0004003/}.

\bibitem[Pikovsky \& Feudel, 1994]{PikovskyFeudel94}
Pikovsky, A. \& Feudel, U. [1994] {``Characterizing strange nonchaotic
  attractors,''} {\em CHAOS} {\bf 5}(1), 253--260.

\bibitem[Stark et~al., 1999]{SFGP99}
Stark, J., Feudel, U., Glendinning, P.  \& Pikovsky, A. [1999] {``Rotation
  numbers for quasiperiodically forced monotone circle maps,''} preprint,
  University College London.

\bibitem[Sturman, 1999]{Sturman99}
Sturman, R. [1999] {``Scaling of intermittent behaviour of a strange nonchaotic
  attractor,''} {\em Phys. Lett. A} {\bf 259}, 355--365.

\end{thebibliography}
}


\begin{appendix}
\section*{Appendix: Details on the Numerical Computations}
\addtocounter{section}{1}

We now explain our numerical computations in more detail and give
values for the accuracy parameters. The Appendix is organized such
that each section is related to one section in the main text. Appendix
\ref{app:sec22} explains how to compute the tongue boundary for the
main tongue with zero rotation number and relates to
Sec.~\ref{sec:app1}. Appendix \ref{app:sec4} discusses the generation
of Fig.~\ref{fig:overlapb} in Sec.~\ref{sec:epsK}. Here, we also
discuss how to identify SNAs using the phase sensitivity exponent,
which is related to the derivative of $x_n$ with respect to
the external phase $\vartheta$. The numerical issues that arise when
determining the nonsmooth pitchfork bifurcation and the accompanying
fractalization process, as reported in Sec.~\ref{sec:nonsmooth},
are described in more detail in App.~\ref{app:sec42}. In contrast to
the method for identifying SNAs in the second region of overlap as
described in App.~\ref{app:sec4}, we used another more efficient
method for determining the regions where SNAs exist in the third
region of overlap. This method is explained in App.~\ref{app:sec43}.

\subsection{Numerical computation of the boundary of the phase-locked
region}
\label{app:sec22}

The boundary $\Omega = \Omega_0(\varepsilon,K)$ of the phase-locked
region with $\rho = 0$ is half its width due to the symmetry. It is 
approximated by estimating the boundary point $\Omega_0$ on a grid of
$320 \times 40$ points in the $(\varepsilon, K)$-plane. Following
\cite{SFGP99} we determine the rotation number~\eqref{eq:rho} within
an accuracy of $\pm 1/N$ by averaging over a sample of 25 orbits of
length $N = F_{28} = 317\,811$ (after 1000 preiterations to eliminate
the effect of transients), where $F_k$ are the Fibonacci numbers $F_1
= F_2 = 1$ and $F_k= F_{k-1} + F_{k-2}$. The Fibonacci numbers $F_k$
are used since the value $\vt_{F_k}$ after $F_k$ iterations is close
to the initial value $\vt_0$ due to the fact that ratios of Fibonacci
numbers are good  rational approximants of our irrational driving
frequency $\omega$. The initial interval $[\Omega_0^-, \Omega_0^+] = [0,
0.2]$ is repeatedly bisected, preserving the relation $\rho(\Omega_0^-)
< 1/N < \rho(\Omega_0^+)$ to ensure that $\Omega_0 \in [\Omega_0^-,
\Omega_0^+]$, until $\Omega_0^+ - \Omega_0^- < \Delta \Omega =
10^{-5}$. Finally, we choose $\Omega_0$ to be the mean value of
$\Omega_0^-$ and $\Omega_0^+$, or zero if the mean value is smaller than
our numerical accuracy $\Delta \Omega$. Note that, as remarked in the
Sec.~\ref{sec:app1}, the question of whether $\Omega_0$ really vanishes or
is just very small cannot be answered by using only numerical methods. 

To distinguish between smooth and nonsmooth saddle-node bifurcations of
invariant curves on the boundary of the phase-locked region we compute the
nontrivial Lyapunov exponent   
\begin{equation}\label{eq:lyapunov}
\lambda(\Omega,\varepsilon,K) 
= \lim_{N\to\infty} \frac1N \sum_{n=0}^{N-1} \ln{\left|\frac{\partial x_{n+1}}{\partial x_n}\right|_{(x_n,\vartheta_n)}} 
= \lim_{N\to\infty} \frac1N \sum_{n=0}^{N-1} \ln{|1+K\cos{2\pi x_n}|} \ .
\end{equation}
Vanishing $\lambda$ (yellow regions in \fig~\ref{fig:maintongue})
indicates smooth saddle-node bifurcation while negative $\lambda$ indicates
nonsmooth saddle-node bifurcation (red regions); see Feudel et
al.~\nocite{FGO97} [1997] and Sec.~\ref{sec:app1}.

\subsection{The cross-section $\Omega = 0$ of the second region of
overlap} 
\label{app:sec4}

Figure~\ref{fig:overlapb} shows the bifurcation structure at a
cross-section $\Omega = 0$. The picture was generated as follows. For 
each grid point $N = F_{32} = 2\,178\,309$ iterations of
Eqs.~\eqref{eq:mapx}--\eqref{eq:maptheta} were computed using 25 
different initial conditions $(x_0, \vt_0 = 0)$. We take advantage of
the fact that inside the tongue with zero rotation number, each
attractor is represented as a single-valued function $x = X(\vt)$,
$\vt \in [0,1)$. (This function is smooth in the case of a nonstrange
attractor and discontinuous everywhere in the case of an SNA.) This
means that the number of different attractors is equal to the number
of different $N$th iterates, i.e., different $x_N$-values. By setting a
tolerance of $\pm 10^{-6}$ the number of attractors was determined
numerically.

To identify the emergence of SNAs, we compute the attractors and
quantify their smoothness properties using the so-called phase
sensitivity exponent introduced by Pikovsky and
Feudel~\nocite{PikovskyFeudel94} [1994]. By formally differentiating
\equ~(\ref{eq:mapx}) with respect to the external phase $\vartheta$,
we get
\begin{equation}
\label{eq:mapderivative}
	\frac{\partial x_{n+1}}{\partial \vartheta} =
	(1+K\cos{2\pi x_n})\frac{\partial x_n}{\partial \vartheta}
			       + 2\pi\varepsilon\cos{2\pi\vartheta_n}.
\end{equation}
When \equ~(\ref{eq:mapderivative}) is iterated together with
Eqs.~(\ref{eq:mapx})--(\ref{eq:maptheta}), starting from some initial
point $(x_0,\vartheta_0)$ and $\partial x_0/\partial \vartheta = 0$,
the phase sensitivity 
\begin{equation}
\label{eq:ps}
	\Gamma_N = \min_{(x_0,\vartheta_0)} \; \max_{0 \leq n \leq N} 
		   \left|\frac{\partial x_n}{\partial \vartheta}\right| 
\end{equation}
diverges like $N^\mu$ for large $N$ in the case of an SNA. On the
other hand, in the case of a smooth attractor it saturates, i.e. the phase
sensitivity exponent $\mu$ is zero. The criteria for saturation
we employ are $\Gamma_N < 10^{15}$ and $\mu < 0.25$ (obtained by
fitting the slope in a ln-ln diagram using three different $N$s). The
black area in Fig.~\ref{fig:overlapb} shows the parameter region for
which $\Gamma_N$ does not saturate when eight different initial points
are iterated for up to $N = F_{33} = 3\,524\,578$ time steps. How much
of this black area persists as $N$ tends to infinity?
Figure~\ref{fig:phase} provides more numerical results for larger $N$
and fixed $K = 0.9$, showing the minimum number of iterations $N =
F_k$ (in terms of the index of the Fibonacci number) for which the
phase sensitivity saturates versus the forcing amplitude
$\varepsilon$. Although a large number of iterations, $F_{43} =
433\,494\,437$, was used, saturation occurs only for $\ve \le 1.566$ and $\ve
\ge 1.568$. Hence the interval $\ve \in [1.566,1.568]$ contains the nonsmooth
pitchfork bifurcation point, as described in Sec.~\ref{sec:nonsmooth}, which
is approximately $1.5676$.
In the next section we discuss the numerical determination of the length of
this gap.  

\subsection{Numerical issues regarding the nonsmooth pitchfork
bifurcation} 
\label{app:sec42}

%
\def\figphase{%
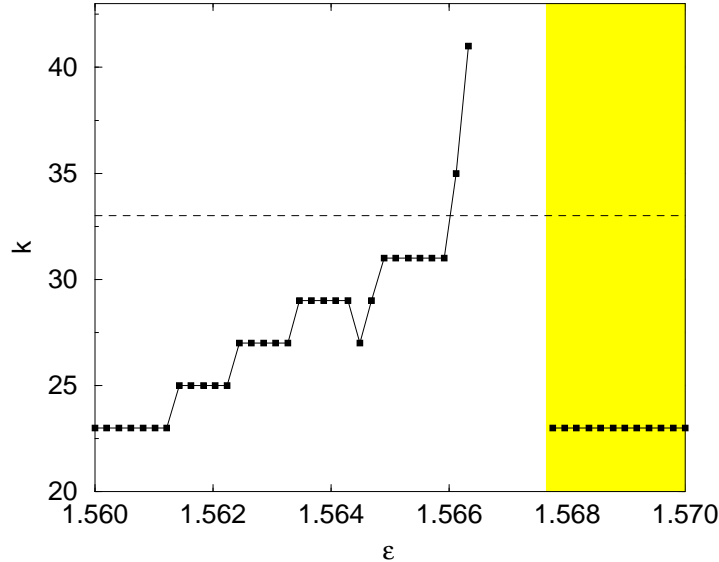
Minimum number of iterations $F_k$ for which the phase sensitivity
saturates, as a function of $\varepsilon$, with $K = 0.9$ and $\Omega =
0$. Compare with \fig~\ref{fig:overlapb}, where $k = 33$ (dashed line)
is used as a threshold. In the yellow marked interval the phase
sensitivity always saturates for $k = 23$, indicating two smooth
attracting invariant curves. In the white marked interval with
$\varepsilon$ smaller than $\approx 1.566$ saturations occurs for $k
\leq 41$, indicating a quite wrinkled but smooth attracting invariant
curve. Using a maximum of $F_{43}$ iterations, there is no saturation
for larger $\varepsilon$.
}
\def\FIGphase{\centerline{\psfig{figure=\figdir/phase.eps,width=9.5cm,angle=0}
}}
\FIGo{fig:phase}{\figphase}{\FIGphase}
Figure~\ref{fig:phase} shows the minimum number of iterations $N =
F_k$ (in terms of the index of the Fibonacci number) for which the
phase sensitivity exponent converges, versus $\ve$. There is no
convergence of the phase sensitivity exponent, even for $N$ as large
as $F_{43} = 433\,494\,437$, in the interval $\ve \in [1.566,
1.568]$. Close to the right side of this interval, a pitchfork 
bifurcation occurs, 
because there is one smooth attractor (white region in
Fig.~\ref{fig:phase}) for smaller $\ve$ and there are two smooth
attractors (yellow region) for larger $\ve$. Hence, we clearly have a
gap when we bound the number of iterations by $F_k = F_{43}$, but it
is unclear whether this gap has nonzero width as $k \to \infty$. 

There are other methods to assess the smoothness properties of the
attractor(s). For example, in \fig~\ref{fig:ratapp} we applied the
method of rational approximations. This method is based on the
approximation of the irrational frequency $\omega$ by rational
frequencies $\omega_k = F_{k-1}/F_k$ with $k \in \N$ and $\omega =
\lim_{k\to\infty} \omega_k$, replacing the quasiperiodically forced
map~(\ref{eq:mapx})--(\ref{eq:maptheta}) with a sequence of
periodically (with period $F_k$) forced maps. The $F_k $th iteration
of such a map is an orientation-preserving diffeomorphism on a circle
depending on $(\Omega,\varepsilon,K)$ and on the initial phase
$\vartheta_0$.  The union of all attracting invariant sets of this 
family of 
diffeomorphisms with $\vartheta_0 \in [0,1/F_k)$, forms the $k$th
approximation of the attractors of the quasiperiodically forced
system. (It is sufficient to consider the subinterval $[0, 1/F_k)$,
since diffeomorphisms with $\vartheta_0 \in [n/F_k,(n+1)/F_k)$, $n =
1,2,\ldots,F_k-1$, are topologically conjugate.) For smooth attractors 
there is a number $k$ for which the rational approximation of order
$k$ and larger does not depend sensitively on
$\vartheta_0$~\cite{PikovskyFeudel94}. 

Figure~\ref{fig:ratapp} shows that this is here the case: the
dependence on the initial phase for moderate $k$ --- one or two stable
fixed points for all $\varepsilon$, depending on $\vartheta_0$ ---
disappears as we cross the lower curve towards higher $k$ values ---
one stable fixed point in the white marked interval, corresponding to
a single smooth attractor in the quasiperiodically forced map, and
two stable fixed points in the yellow marked interval, corresponding to
a pair of smooth attractors. 

However, as mentioned in~\cite{PikovskyFeudel94}, a vanishing
dependence on $\vartheta_0$ is a necessary but not sufficient
condition for smoothness. We also have to stipulate that the maximum
derivative of the attracting sets with respect to $\vartheta_0$ is
bounded for all $\vartheta_0 \in [0,1)$ as $k\to\infty$. Note that it
is nevertheless sufficient to determine the attracting sets only in
the subinterval $[0,1/F_k)$ since the other parts can be obtained by
iterating the map~(\ref{eq:mapx})--(\ref{eq:maptheta}) with the
rational frequency $\omega_k$. Furthermore, it is elegant to iterate
\equ~(\ref{eq:mapderivative}) simultaneously in order to determine the 
derivative. The maximum derivative obtained by this procedure is an
approximation of the phase sensitivity. Using the same criteria for
boundedness as before, we surprisingly find that the maximum
derivative always saturates at some order $k$, shown by the upper
curve in \fig~\ref{fig:ratapp}. The qualitative features of
\figs~\ref{fig:phase}--\ref{fig:ratapp} persist if we choose
different $K$s in the region with large phase sensitivity.  

\def\figratapp{%
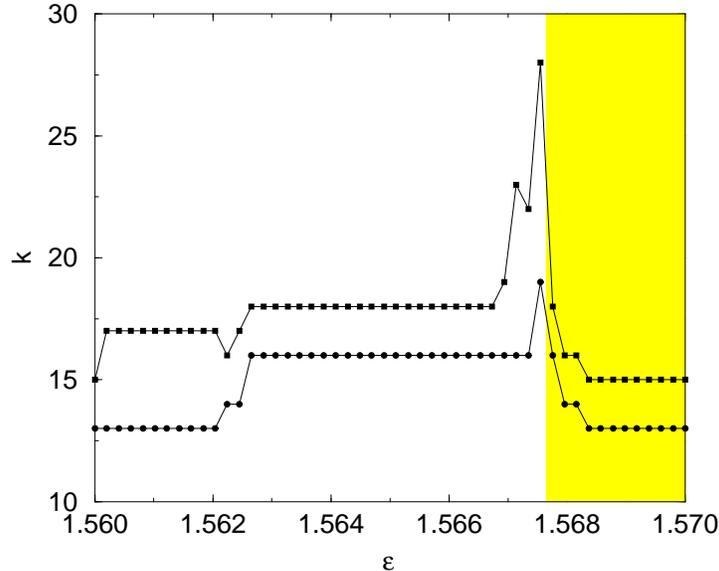
Minimum order $k$ of the rational approximation for which the sensitive
dependence on the initial phase $\vartheta_0$ vanishes (lower curve)
and for which the maximum derivative of the attracting sets in
$\vartheta_0 \in [0,1)$ saturates (upper curve), as a function of
$\varepsilon$ with $K = 0.9$ and $\Omega = 0$,
cf. \fig~\ref{fig:phase}. We used 1000 different values of
$\vartheta_0 \in  [0,1/F_k)$.
} 
\def\FIGratapp{%
\centerline{\psfig{figure=\figdir/ratapp.eps,width=9.5cm,angle=0}
}}
\FIGo{fig:ratapp}{\figratapp}{\FIGratapp}

\subsection{SNAs in the third region of overlap}
\label{app:sec43}

To compute SNAs near the boundary of the tongue in the third region of
overlap we use a more effective method than the method described in
App.~\ref{app:sec4}. This method takes advantage of the fact that the
corresponding SNAs are unbounded in the $x$-direction in the lift of
the map. We define the amplitude of an attractor at time
$N$ as
\begin{displaymath}
	\tilde{\Gamma}_N = \min_{(x_0,\vartheta_0)} 
			\left(\max_{0 \leq n \leq N} x_n
				-\min_{0 \leq n \leq N} x_n \right).
\end{displaymath}
For an unbounded SNA the asymptotic behavior of the amplitude is given
by $c\ln{N}$~\cite{FKP95}, in all other cases the amplitude saturates
for large $N$. The numerical algorithm is almost the same as for the
phase sensitivity~(\ref{eq:ps}). However, by fitting the slope using
four different $N$, we find as criteria for saturation
$\tilde{\Gamma}_N < 6.0$ and $c < 0.02$. Therefore, typically far less 
iterations are necessary, so that a larger maximum number of
iterations $N = F_{40} = 102\,334\,155$ and 10 different initial
conditions can be used.  

\end{appendix}

\end{document}